\documentclass[12pt,preprint]{aastex}

\usepackage[tbtags,fleqn]{amsmath}
\usepackage{amsfonts}
\usepackage{pstricks, pst-plot}
\usepackage{graphicx}
\usepackage{natbib}

\newcommand{\e}{\mathrm e}
\newcommand{\diff}{\mathrm d}

\newcommand{\mincir}{\raise
  -2.truept\hbox{\rlap{\hbox{$\sim$}}\raise5.truept \hbox{$<$}\ }}
\newcommand{\magcir}{\raise
  -2.truept\hbox{\rlap{\hbox{$\sim$}}\raise5.truept \hbox{$>$}\ }}
\newcommand{\Cluster}{RDCS~1252.9$-$2927}
\newcommand{\cluster}{RDCS1252}


\begin{document}

\title{HST/ACS weak lensing analysis of the galaxy
  cluster \Cluster\ at $z=1.24$ \altaffilmark{0} }

\author{M. Lombardi\altaffilmark{1,2},
  P. Rosati\altaffilmark{1}, J.P. Blakeslee\altaffilmark{3},
  S. Ettori\altaffilmark{1,8}, R. Demarco\altaffilmark{3},
  H.C. Ford\altaffilmark{3}, G.D. Illingworth\altaffilmark{5}, 
  M. Clampin\altaffilmark{7}, G.F. Hartig\altaffilmark{7},
  N. Ben\'\i tez\altaffilmark{3}, T.J. Broadhurst\altaffilmark{4}, 
  M. Franx\altaffilmark{6}, M.J. Jee\altaffilmark{3},
  M. Postman\altaffilmark{7}, R.L. White\altaffilmark{7}}
\altaffiltext{1}{%
  European Southern Observatory,
  D-85748 Garching bei M\"unchen, Germany.}
\altaffiltext{2}{%
  University of Milan, Department of Physics, via Celoria 16, I-20133
  Milan, Italy.}
\altaffiltext{3}{%
  Department of Physics and Astronomy, Johns Hopkins University,
  Charles and 34th Street, Bloomberg Center, Baltimore, MD 21218.}
\altaffiltext{4}{%
  Racah Institute of Physics, Hebrew University, Jerusalem 91904,
  Israel.}
\altaffiltext{5}{%
  UCO/Lick Observatory, University of California, Santa Cruz, CA
  95064.}
\altaffiltext{6}{%
  Leiden Observatory, Postbus 9513, 2300 RA Leiden, Netherlands.}
\altaffiltext{7}{%
  STScI, 3700 San Martin Drive, Baltimore, MD 21218.}
\altaffiltext{8}{%
  INAF, Osservatorio Astronomico di Bologna, via Ranzani 1, I-40127
  Bologna, Italy}
\altaffiltext{0}{Based in part on observations obtained at the
  European Southern Observatory using the ESO Very Large Telescope on
  Cerro Paranal (ESO program 166.A-0701).}

\authoremail{mlombard@eso.org}
  
\begin{abstract}
  We present a weak lensing analysis of one of the most distant
  massive galaxy cluster known, \Cluster\ at $z = 1.24$, using deep
  images from the Advanced Camera for Survey (ACS) on board the Hubble
  Space Telescope (HST).  By taking advantage of the depth and of the
  angular resolution of the ACS images, we detect for the first time
  at $z > 1$ a clear weak lensing signal in both the $i$ (F775W) and
  $z$ (F850LP) filters.  We measure a $5$-$\sigma$ signal in the $i$
  band and a $3$-$\sigma$ signal in the shallower $z$ band image.  The
  two radial mass profiles are found to be in very good agreement with
  each other, and provide a measurement of the total mass of the
  cluster inside a $1 \mbox{ Mpc}$ radius of $M(< 1 \mbox{ Mpc}) =
  (7.3 \pm 1.3) \times 10^{14} \, \mbox{M}_\odot$ in the current
  cosmological concordance model $h = 0.70$, $\Omega_\mathrm{m} =
  0.3$, $\Omega_\Lambda = 0.7$, assuming a redshift distribution of
  background galaxies as inferred from the Hubble Deep Fields surveys.
  A weak lensing signal is detected out to the boundary of our field
  ($3 \arcmin$ radius, corresponding to $1.5 \mbox{ Mpc}$ at the
  cluster redshift).  We detect a small offset between the centroid of
  the weak lensing mass map and the brightest cluster galaxy, and we
  discuss the possible origin of this discrepancy.  The cumulative
  weak lensing radial mass profile is found to be in good agreement
  with the X-ray mass estimate based on \textit{Chandra\/} and
  \textit{XMM-Newton\/} observations, at least out to $R_{500} \simeq
  0.5 \mbox{ Mpc}$.
\end{abstract}

%


\section{Introduction}
\label{sec:introduction}

Cluster of galaxies lie at the extreme of the mass spectrum of
gravitationally bound structures, and therefore their physical properties
are thought to be mainly driven by gravitational processes.  
The study of the mass distribution of galaxy clusters conveys precious
information on the relationship between dark and luminous matter and
can be used to test cosmological models
\citep[e.g.][]{1996MNRAS.282..263E,1998ApJ...504....1B}.

Over the last decade, gravitational lensing has proved to be a powerful
method to determine the mass and the mass distribution of galaxy
clusters (see \citealp{RevBS} for a review).  The weak lensing
technique, based on a statistical analysis of small distortions of
faint background sources, is a particularly valuable tool since it
relies on simple, well verified assumptions and is sensitive to the
\textit{total\/} mass of a cluster, regardless of its physical
state and spatial distribution.

Following the early work of \citet*{1990ApJ...349L...1T}, many massive
clusters have been the subject of weak lensing analyses
\citep[e.g.][]{1996A&A...314..707S, 1996ApJ...469...73S,
  2000A&A...363..401L}.  In most cases, these studies focused on many
clusters at low-to-medium redshifts ($0.1<z<0.5$), for which
moderately deep imaging is sufficient to successfully apply this
technique.  The first attempt to study the weak lensing signal of a
high-redshift cluster (Cl~1604$+$4304 at $z = 0.89$) was not
successful (\citealp{1994MNRAS.270..245S}; but see
\citealp{2005AJ....129...20M} for a recent weak lensing detection).
The velocity dispersion of this cluster was initially estimated to be
$1226^{+245}_{-154} \mbox{ km s}^{-1}$ \citep{1998AJ....116..560P},
however a recent extensive spectroscopic study
\citep{2004ApJ...607L...1G} has found a clear evidence for a
superposition of four moderate mass systems (with velocity dispersions
$\sigma_v \lesssim 800 \mbox{ km s}^{-1}$) within $\Delta z\simeq
0.1$.  \citet{1997ApJ...475...20L} reported the first clear detection
of a weak-lensing signal from a distant cluster, MS~1054.4$-$0321 at
$z = 0.83$, using deep ground-based images.  An improved analysis of
MS1054 was presented by \citealp{2000ApJ...532...88H}) using HST/WFPC2
observations, which showed the power of weak lensing studies with HST
when point spread function (PSF) effects are properly taken into
account. Weak lensing analysis of very luminous X-ray clusters at $z
\simeq 0.8$ is by now relatively straightforward (e.g.
\citealp{1998ApJ...497L..61C} studied MS~1137.5+6625 at $z = 0.783$
and RXJ~1716.4$+$6708 at $z = 0.809$; \citealp{2004AJ....127.1263H}
studied RXJ~0152.7$-$1357 at $z=0.83$).

Weak lensing studies of clusters at $z \gtrsim 1$ are instead
particularly challenging and essentially impossible with ground-based
observations.  Most of the galaxies observed even in the deepest
ground-based observations have redshifts smaller than unity, and thus
are foreground with respect to high-redshift clusters, or are only
weakly lensed.  To avoid a severe dilution of the lensing signal one
needs to accurately select background galaxies, typically using
photometric redshifts.  In addition, galaxies at redshifts larger then
unity are faint and small, so that measurements of their
ellipticities, needed for the weak lensing analysis, are difficult
because of photometric uncertainties and the smearing effect of the
PSF (in addition to the seeing in ground-based images). Finally,
because of the redshift dependence of the lensing signal, weak lensing
masses of distant clusters are strongly sensitive to the redshift
distribution of the background galaxies.  For example, it is
relatively easy to show that a systematic relative error of $5\%$ on
$(1 + z)$ of the background galaxies produces a systematic error of
$15\%$ on the lensing mass estimate of a $z = 1$ galaxy cluster.

The advent of the Advanced Camera for Surveys (ACS) on board of the
HST has given a unique opportunity to study the mass distribution of
clusters at redshift larger than unity via weak lensing techniques.
The combination of the excellent PSF and the much improved CCD
sensitivity overcomes many of the challenges posed by distant
clusters.  In this paper, we report an unambiguous weak lensing
detection of the galaxy cluster \Cluster\ at redshift $z = 1.237$
(hereafter \cluster\ for brevity) using a mosaic of four ACS pointings
in the $i$ and $z$ bands, the first of this kind at $z>1$.

The paper is organized as follow.  In Sect.~\ref{sec:observations} we
describe the observations and data reduction.  The weak lensing
analysis is presented in Sect.~\ref{sec:weak-lens-analys} and the
results obtained are discussed in Sect.~\ref{sec:results}.  In
Sect.~\ref{sec:discussion} we summarize our conclusions.  Finally,
App.~\ref{sec:mass-apert-stat} briefly reports on the mass aperture
statistics.

We adopt the current ``concordance'' cosmological model: $H_0 = 70
\mbox{ km s}^{-1} \mbox{ Mpc}^{-1}$, $\Omega_\mathrm{m} = 0.3$, and
$\Omega_\Lambda = 0.7$.  In this model, one arcminute at the cluster
redshift corresponds to a linear size of $0.5 \mbox{ Mpc}$.

\section{Observations}
\label{sec:observations}

\cluster\ was originally discovered as an extended X-ray source in the
\textit{ROSAT\/} Deep Cluster Survey (RDCS;
\citealp{1998ApJ...492L..21R}) and since then has been the target of a
large number of follow-up observations, which include a VLT Large
Program with FORS2 optical imaging and spectroscopy and ISAAC deep
near infrared imaging \citep{2004A&A...416..829L}. 

\cluster\ was observed in the F775W and F850LP bandpasses in May 2002
and June 2002 with the HST/ACS Wide Field Camera as part of the
Guaranteed Time Observation program (proposal 9290).  The observations
were done in a $2 \times 2$ mosaic pattern, with 3 and 5 orbits of
integration in F775W and F850LP, respectively, at each of the four
pointings.  Two exposures were taken per orbit, and we dithered by
2~pixels in both the $x$ and $y$ directions between orbits.  However,
the imaging was split between two `epochs' separated in time by about
six weeks, and there was a $\sim 20 \mbox{ pixels}$ ($1''$) offset in
the pointing between the two epochs as a result of differences in the
guide star acquisition.  The first three orbits of F850LP imaging for
each position was done during the first epoch (early/mid May), while
the remaining two orbits in F850LP and all three orbits in F775W were
done during the second epoch (mid/late June).

These data have been described by \citet{2003ApJ...596L.143B}, who
processed the images as a single large mosaic using \textit{Apsis\/}
(described by \citealp{2003adass..12..257B}) and studied the
color-magnitude relation of cluster galaxies.  In order to improve the
modeling of the point spread function (PSF) variations, we processed
each of the four pointing separately with \textit{Apsis\/} and used an
output scale of $0.025 \mbox{ arcsec pixel}^{-1}$ to achieve exquisite
PSF sampling.  We also used the Gaussian \textit{drizzle\/}
\citep{2002PASP..114..144F} interpolation kernel because the
resolution-preserving quality of the Lanczos kernel used by
\citet{2003ApJ...596L.143B} is not necessary in case of oversampling.
We calibrate our photometry to the AB system using zero points of
25.654 and 24.862 for F775W and F850LP, respectively (Sirianni et al.\ 
2004, in preparation) and adopt a Galactic reddening for this field of
$E(i{-}z) = 0.041$ mag based on the \citet{1998ApJ...500..525S} dust
maps.

The cluster was also the object of deep observations with
\textit{Chandra\/} and \textit{XMM-Newton\/} (see
\citealp{2004AJ....127..230R}).  These data have provided a fairly
accurate temperature and metallicity of the intra-cluster gas, $k T =
(6.5 \pm 0.5) \mbox{ keV}$ and $Z = 0.49^{+0.08}_{-0.13} \ Z_\odot$
(if the \textit{Chandra\/} data alone are used, we obtain $k T =
6.5^{+0.5}_{-0.9} \mbox{ keV}$ and $Z = (0.60 \pm 0.22) \ 
Z_\odot$).\footnote{The X-ray data analysis in this paper has been
  here revised with respect to \citet{2004AJ....127..230R} and
  \citet{2004A&A...417...13E} by adopting new calibration files
  (version 2.28), with a proper CTI, time dependent gain corrections,
  and VFAINT cleaning applied to the event-one file.  Moreover, a
  reverse edge at $2.07 \mbox{ keV}$ is added to the thermal spectrum
  model as suggested by A.~Vikhlinin (private communication) to
  account for contamination of the CCD by methylene.  As a result, the
  measured temperature increases by 1--2~$\sigma$ with respect to the
  results obtained with previous (version 2.21) calibration files.}\@
Thanks to its high angular resolution, \textit{Chandra\/} has also
revealed some departure from a spherically symmetric distribution of
the gas (see below).

\section{Weak lensing analysis}
\label{sec:weak-lens-analys}

\subsection{Basic relations}
\label{sec:basic-relations}

In this subsection we briefly review the basic relations used in weak
lensing mass reconstructions.  We mainly use the notation of
\citet{RevBS}.

For a gravitational lens at redshift $z_\mathrm{d}$ (in our case
$z_\mathrm{d} = 1.237$) and a source at redshift $z > z_\mathrm{d}$, we
define the critical density $\Sigma_\mathrm{c}(z)$ as
\begin{equation}
  \label{eq:1}
  \Sigma_\mathrm{c}(z) = \frac{c^2}{4 \pi G} \frac{D(z)}{D(z_\mathrm{d})
  D(z_\mathrm{d}, z)} \; ,
\end{equation}
where $D(z) = D(0,z)$ and $D(z_1, z_2)$ is the angular diameter
distance between objects at redshift $z_1$ and redshift $z_2$.  A lens
with projected mass density larger than $\Sigma_\mathrm{c}(z)$ in its
core can produce strong lensing effects such as multiple images;
instead, a lens with $\Sigma \ll \Sigma_\mathrm{c}(z)$ for any $z$
only produces weak effects, detectable through a statistical analysis.
In the following we will focus on the analysis in the weak lensing
regime, while we defer the study of strong lensing effects to another
paper.

Accurate measurements of the shapes (ellipticities) of background
galaxies lead to the estimate of the lens (reduced) \textit{shear\/}
$g(\vec \theta)$ for any angular direction $\vec\theta$.  These shear
maps will inevitably have a limited resolution which is basically set
by the density of background galaxies
\citep[see][]{1998A&A...335....1L}.  The reduced shear that acts on a
given galaxy depends on the galaxy redshift; however, a weak lensing
analysis can still be carried out if we know the \textit{redshift
  distribution\/} $p(z)$ of the background galaxies (without knowing
necessarily the individual redshifts; see, e.g., \citealp{SS3}).  For
practical purposes, we can perform the weak lensing analysis
\textit{as if\/} all background galaxies were at the same redshift
$z_\mathrm{eff}$ [one can easily show that in the weak lensing limit
this is a legitimate simplification; see Eq.~\eqref{eq:6} below].  In
the weak lensing limit, the shear can be directly inverted into the
lens \textit{convergence\/} $\kappa(\vec\theta) = \Sigma(\vec\theta) /
\Sigma_\mathrm{c}$ \textit{up to an arbitrary additive constant\/}; in
other words, the transformation $\kappa \mapsto \kappa' = \kappa +
\lambda$ leaves all observables unchanged.  This is referred as
``mass-sheet degeneracy'' and is often broken by fitting the mass
profile with a parametric model (and testing to what extent this
depends on the model).

\subsection{Ellipticity and shear measurements}
\label{sec:ellipt-meas}

The lensing analysis was carried out using the \textsc{Imcat} software
\citep{1995ApJ...449..460K} with some significant modifications
\citep[see][]{1997ApJ...475...20L, 2001A&A...366..717E}.  The whole
lensing reduction pipeline, which we describe in detail below, was
carefully tested using synthetic field images generated with the
\textit{Skymaker\/} program \citep{2001A&A...366..717E}.

For each of the four ACS pointings, we combined the $i$ and $z$ bands
and made a single master catalog.  This catalog was used to perform
the object detection with the \textsc{Imcat} hierarchical peak finding
algorithm (\texttt{hfindpeaks}), using a set of Gaussian kernels with
radii in the range $r_\mathrm{g} \in [0.5, 50] \mbox{ pixels}$.  The
catalog was then visually inspected and spurious detections (such as
star spikes, objects close to very bright sources, or near the field
edges) were eliminated.


The rest of the analysis was then performed on the $i$ and $z$ bands
\textit{independently}.  We measured the local sky and its gradient
around each object by computing the mode of pixel values on 4 annular
sectors (with internal/external radii set to $3$ and $6$ times the
detection radius $r_\mathrm{g}$) using the \textsc{Imcat} utility
\texttt{getsky}.  We then performed aperture photometry
(\texttt{apphot}) and shape measurements (\texttt{getshapes}) for each
object.  Along these steps we explicitly removed closed pairs (i.e.\ 
objects whose distance was smaller than $3 r_\mathrm{g}$) and objects
with negative quadrupole moments.

We then classified objects as stars, galaxies, or spurious sources on
the magnitude vs.\ half-light radius $r_\mathrm{h}$ plot.  Unsaturated
stars occupy a very narrow region on this plot characterized by small
$r_\mathrm{h}$, while well detected galaxies have larger radii; faint
objects with sizes comparable to the size of the point spread function
were discarded because no clear identification was possible.

\begin{figure*}[!t]
  \begin{center}
    \includegraphics[bb=157 271 460 575, width=0.48\hsize,
    keepaspectratio]{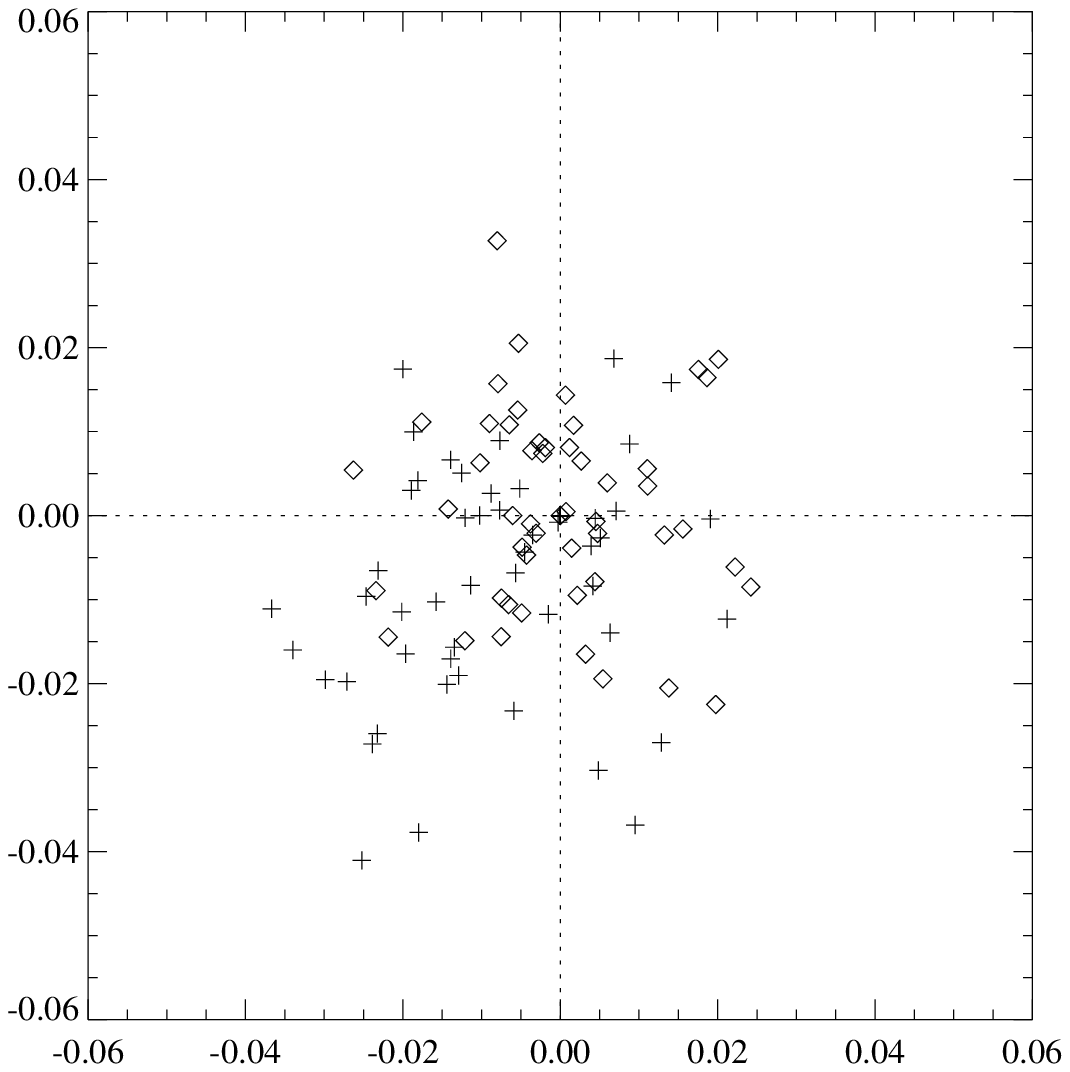}
    \hfill
    \includegraphics[bb=157 271 460 575, width=0.48\hsize,
    keepaspectratio]{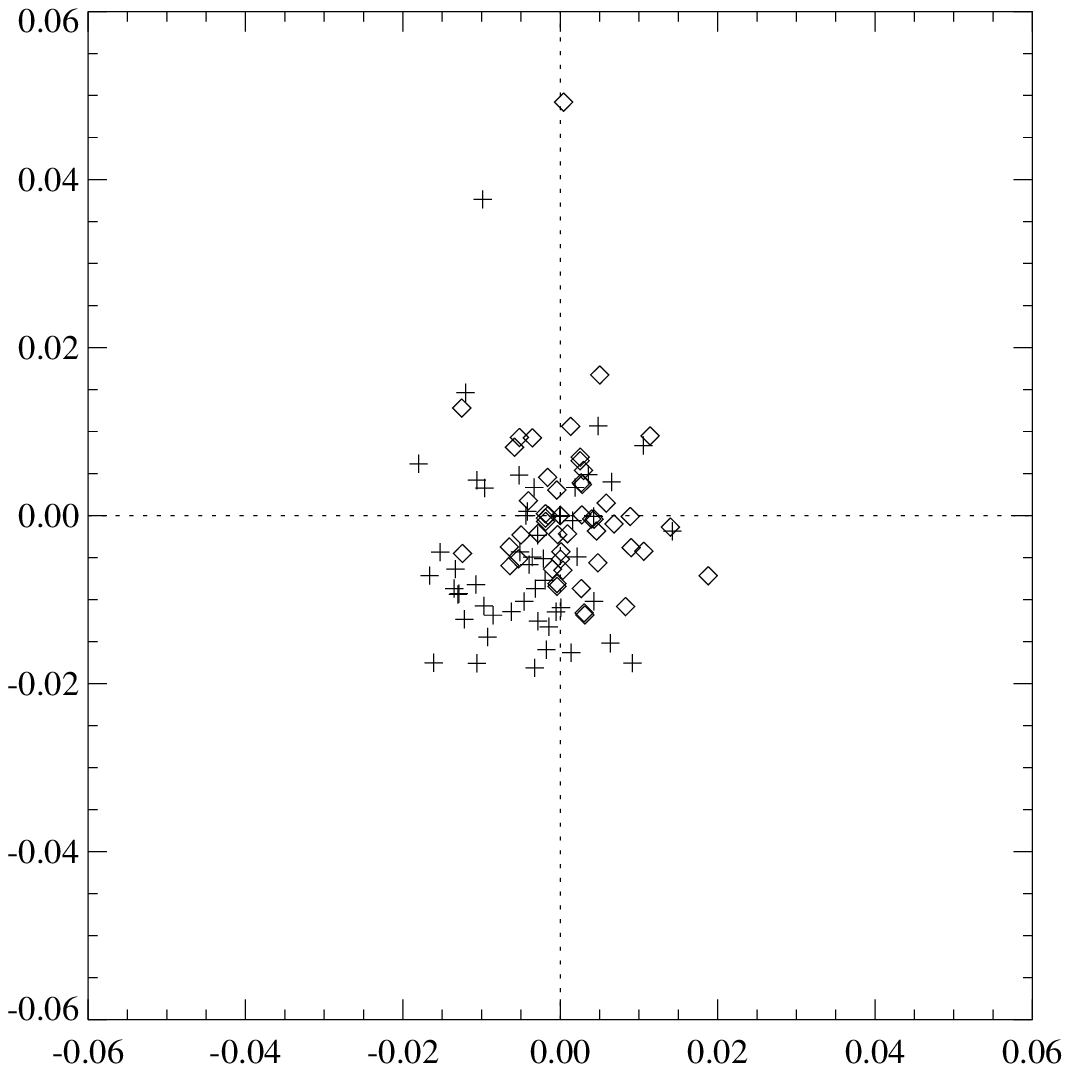}
    \caption{The distribution on the complex plane of the star
      ellipticities $\chi$ in the $i$ (left) and $z$ (right) band.
      The original, uncorrected ellipticities are marked with crosses,
      and the corrected ones with diamonds.  After the correction, we
      obtain $\bigl\langle \lvert \chi \rvert^2 \bigr\rangle^{1/2}
      \simeq 0.011$ in $i$ band and $\bigl\langle \lvert \chi \rvert^2
      \bigr\rangle^{1/2} \simeq 0.008$ in $z$ band.}
    \label{fig:1}
  \end{center}
\end{figure*}

\begin{figure*}[!t]
  \begin{center}
    \includegraphics[bb=127 249 468 591, width=0.48\hsize, 
    keepaspectratio]{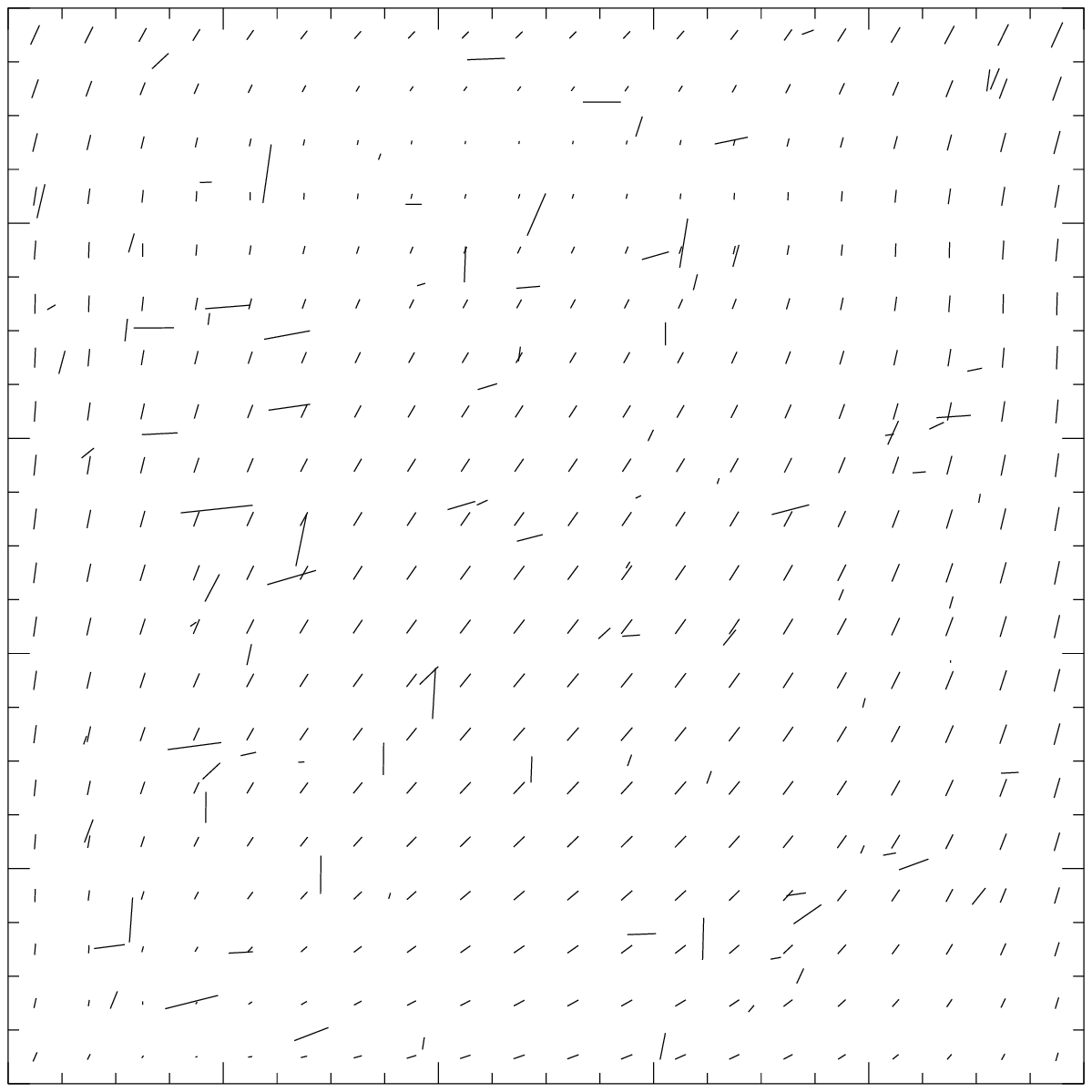}
    \hfill
    \includegraphics[bb=127 249 468 591, width=0.48\hsize,
    keepaspectratio]{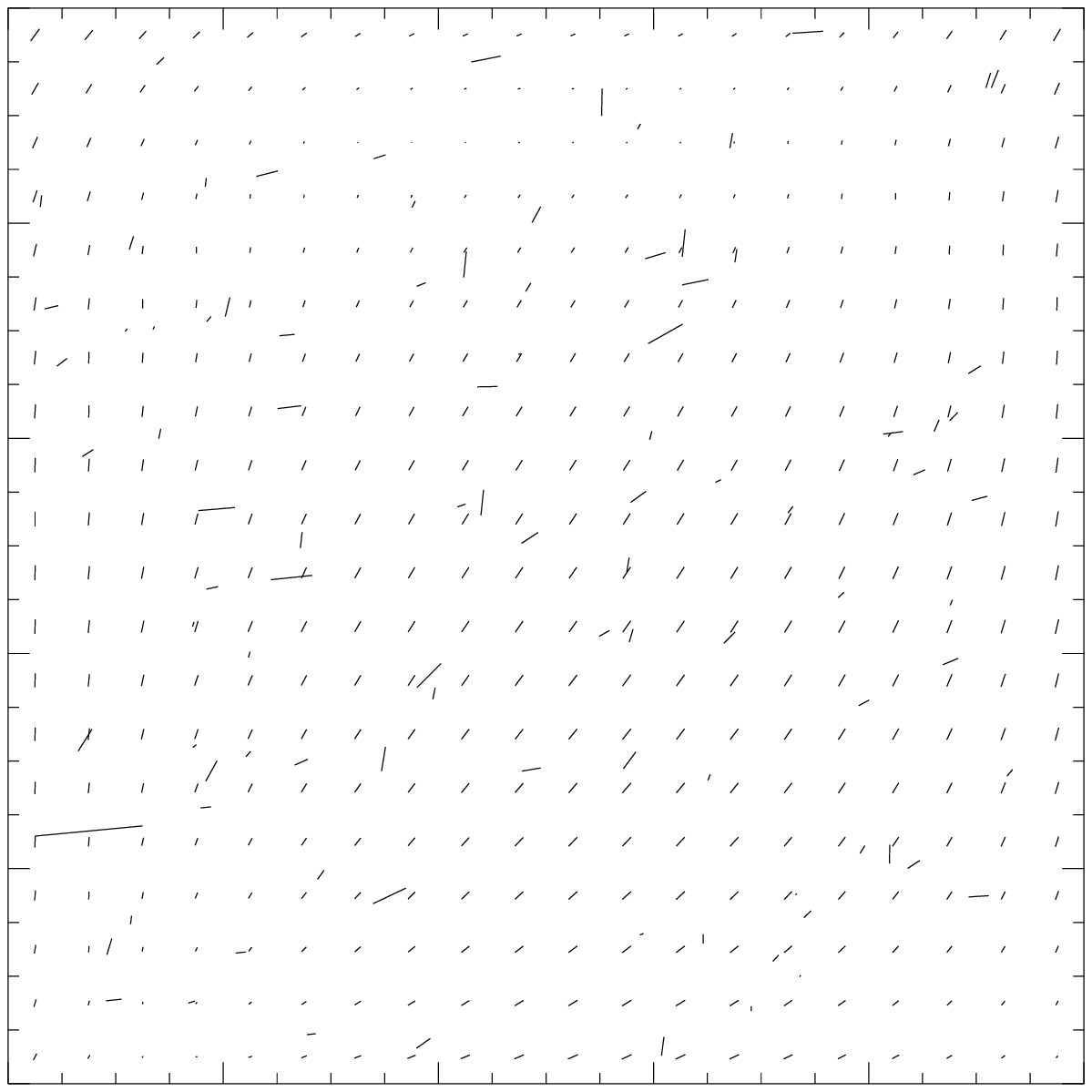}
    \caption{The ellipticity pattern in the $i$ (left) and $z$ (right)
      band.  We plot the fitted ellipticities (a second order
      polynomial) on a regular grid ; the original star ellipticities
      are also marked in this figure at the star locations.  The
      longest tick are associated with ellipticities of $\sim 3\%$ on
      the left plot, and $\sim 5\%$ on the right plot.}
    \label{fig:2}
  \end{center}
\end{figure*}

For both galaxies and stars we thereby measured the complex
ellipticity $\chi$.  As described in \citet{1995ApJ...449..460K}
\citep[see also][]{1997ApJ...475...20L}, the observed ellipticity of
an object is related to the true, unlensed ellipticity $\chi^0$ (in
fact, the unlensed ellipticity convolved with an isotropic kernel; see
\citealp{RevBS} for details) through
\begin{equation}
  \label{eq:2}
  \chi - \chi^0 = P^\mathrm{g} g - P^\mathrm{sm} q \; ,
\end{equation}
where $g$ is the complex shear, $q$ is a quantity representing the
anisotropic part of the PSF, $P^\mathrm{sm}$ is the \textit{smear
  polarizability}, and $P^\mathrm{g}$ is given by
\begin{equation}
  \label{eq:3}
  P^\mathrm{g} = P^\mathrm{sh} - P^\mathrm{sm} \bigl(
  P^\mathrm{sm*}\bigr)^{-1} P^\mathrm{sh*} \; .
\end{equation}
Here $P^\mathrm{sh}$ is the \textit{shear polarizability}, and stars
($*$) as superscript denote the corresponding quantities evaluated for
stellar objects.

Stars were used to measure the anisotropy of the PSF, characterized by
$q$, and to calibrate the ellipticity-shear relation, represented by
$P^\mathrm{g}$.  Given the relatively low galactic latitude ($b =
+33^\circ$) of \cluster, more than 300 unsaturated stars are available
across the ACS field.  Figure~\ref{fig:1} shows the observed star
ellipticities $\chi$ in the $i$ and $z$ bands on the complex plane.
The original, uncorrected star ellipticities were in most cases
already very small (RMS below $2\%$); however, by fitting a
second-order polynomial on the field, we obtained star ellipticity
residuals as small as $\bigl\langle | \chi |^2 \bigr\rangle^{1/2}
\simeq 0.011$ for the $i$ band and $\bigl\langle | \chi |^2
\bigr\rangle^{1/2} \simeq 0.008$ for the $z$ band (this smaller value
obtained in the $z$ band is explained by observing that the size of
the isotropic part of the PSF is larger in $z$ than in $i$).  Note
that, for each band, the fitting was performed independently on the
four ACS pointings.  We also verified that the use of individual fits
for each of the two ACS chips did not significantly decrease the
residuals on the corrected star ellipticities.  Figure~\ref{fig:2}
shows the observed patterns on $i$ and $z$ for one of the pointings;
the other pointings show consistently similar patterns.

The ellipticity-shear relation was calibrated by measuring the
quantity $\bigl( P^\mathrm{sm*}\bigr)^{-1} P^\mathrm{sh*}$ for star
objects.  Following \citet{2001A&A...366..717E}, we evaluated this
quantity for each star using different filter scales (with
$r_\mathrm{g}$ ranging from $2$ to $10$ pixels).  For any scale, then,
we took the average of this quantity evaluated on all stars.  Finally,
when calculating $P^\mathrm{g}$ for a given galaxy, we used the scale
corresponding to the $r_\mathrm{g}$ of that galaxy.

\begin{figure*}[t]
  \centering
  \includegraphics[bb=144 258 458 556, width=0.48\hsize,
  keepaspectratio]{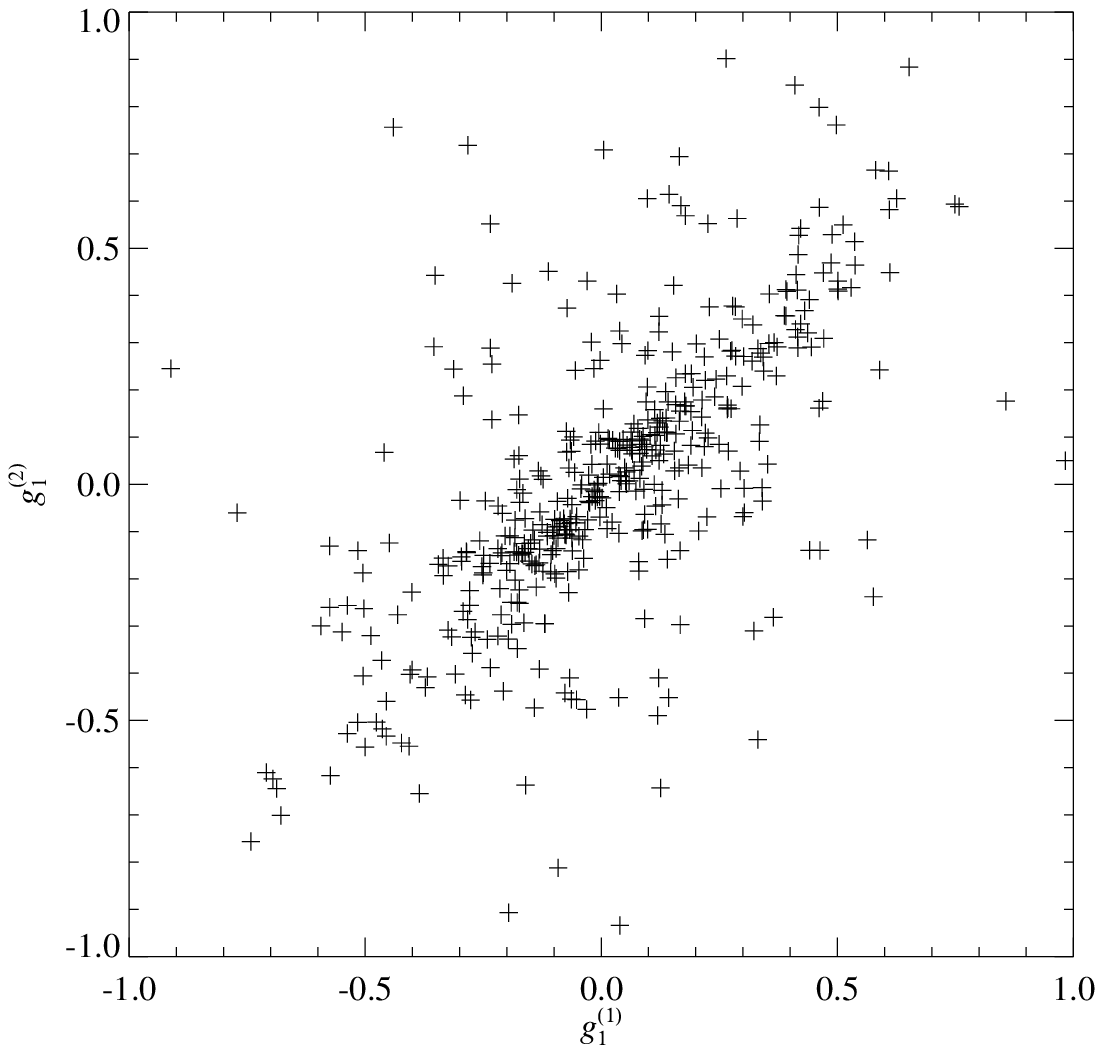}
  \hfill
  \includegraphics[bb=144 258 458 556, width=0.48\hsize,
  keepaspectratio]{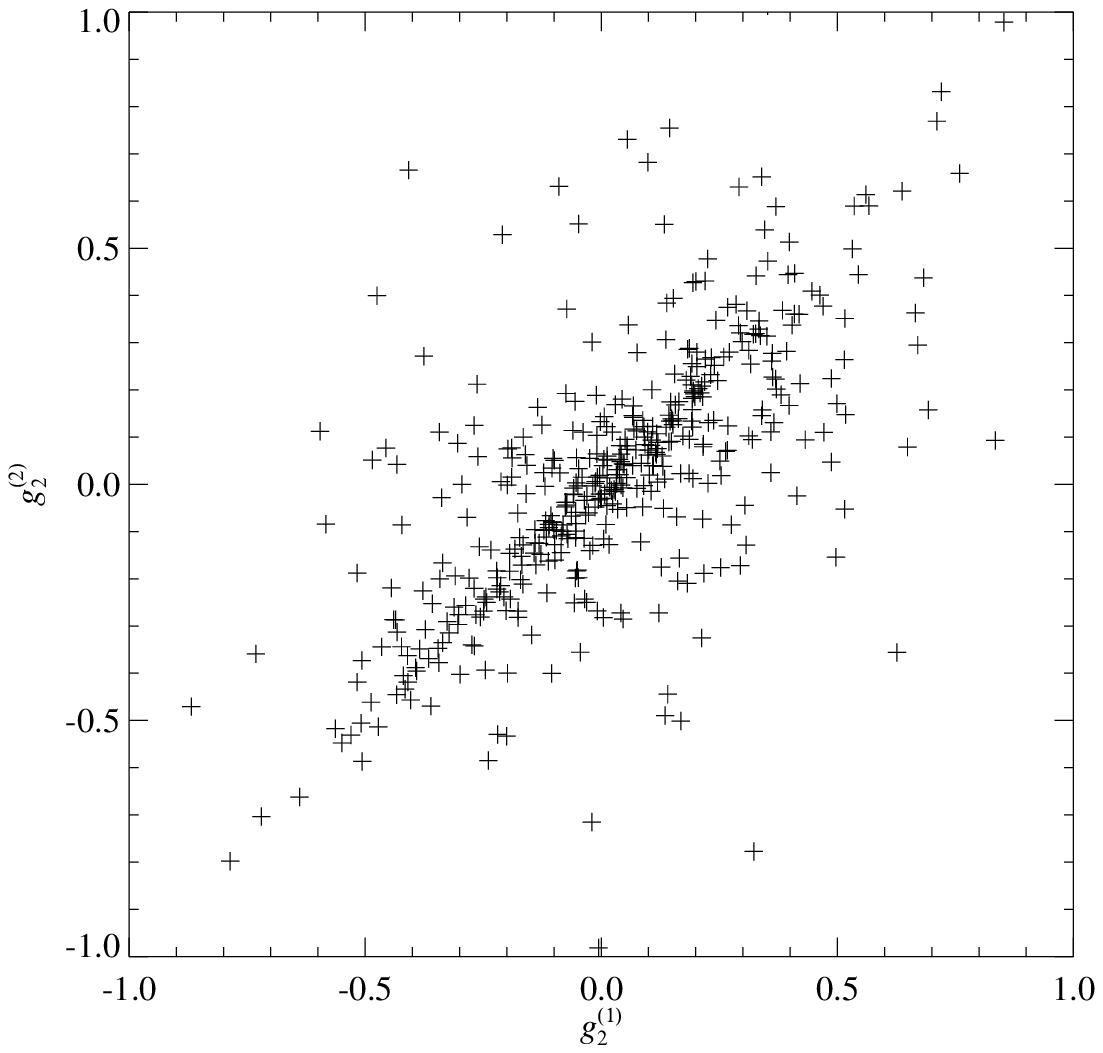}
  \caption{Measured galaxy shears of galaxies observed in two
    adjacent pointings of the $i$ band mosaic.  The left (respectively
    right) plot shows the real (imaginary) component of the shear
    measured in the second pointing versus the shear measured in the
    first one.}
  \label{fig:3}
\end{figure*}

\begin{figure*}[t]
  \centering
  \includegraphics[bb=144 258 458 556, height=0.48\hsize,
  keepaspectratio]{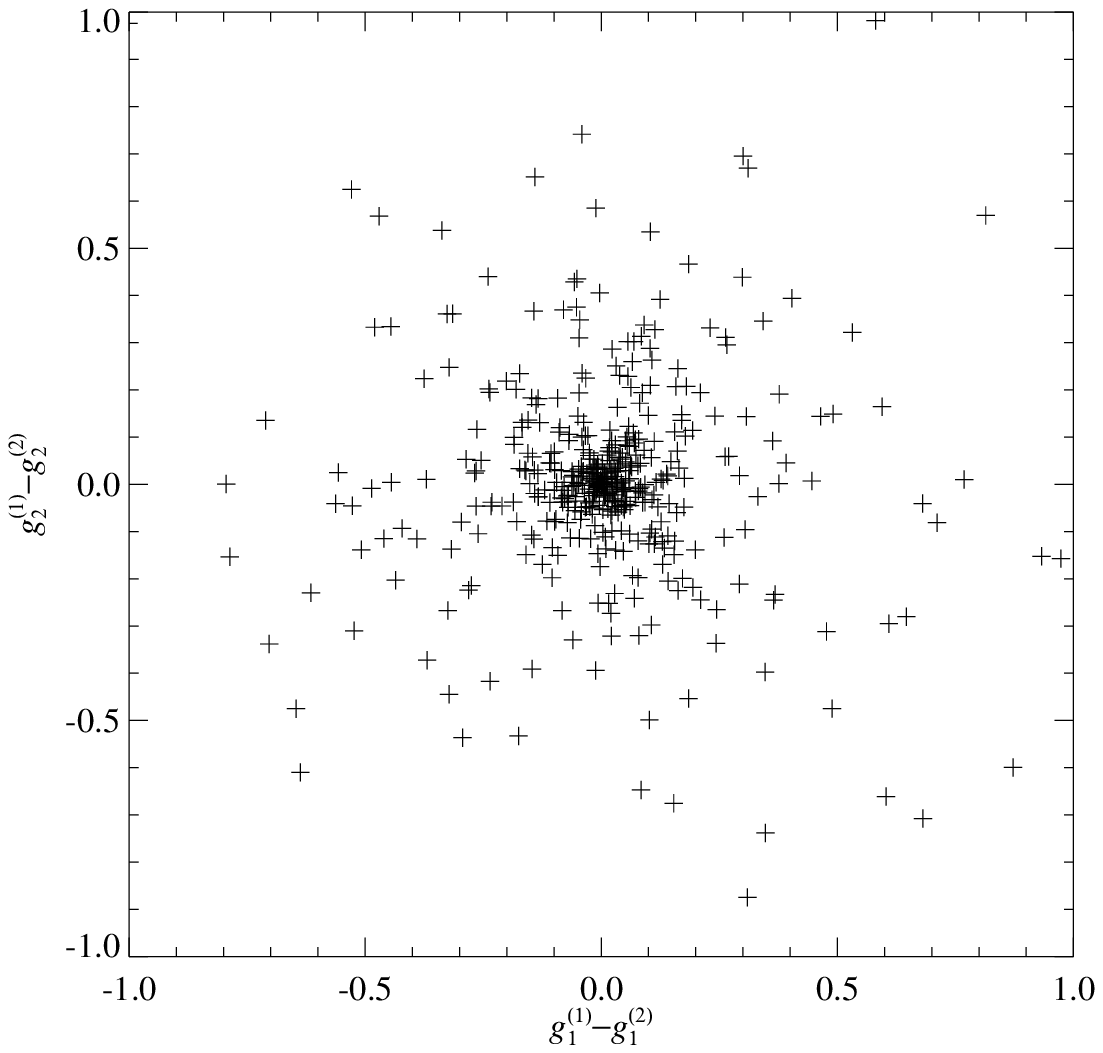}
  \hfill
  \includegraphics[bb=147 259 464 575, height=0.48\hsize,
  keepaspectratio]{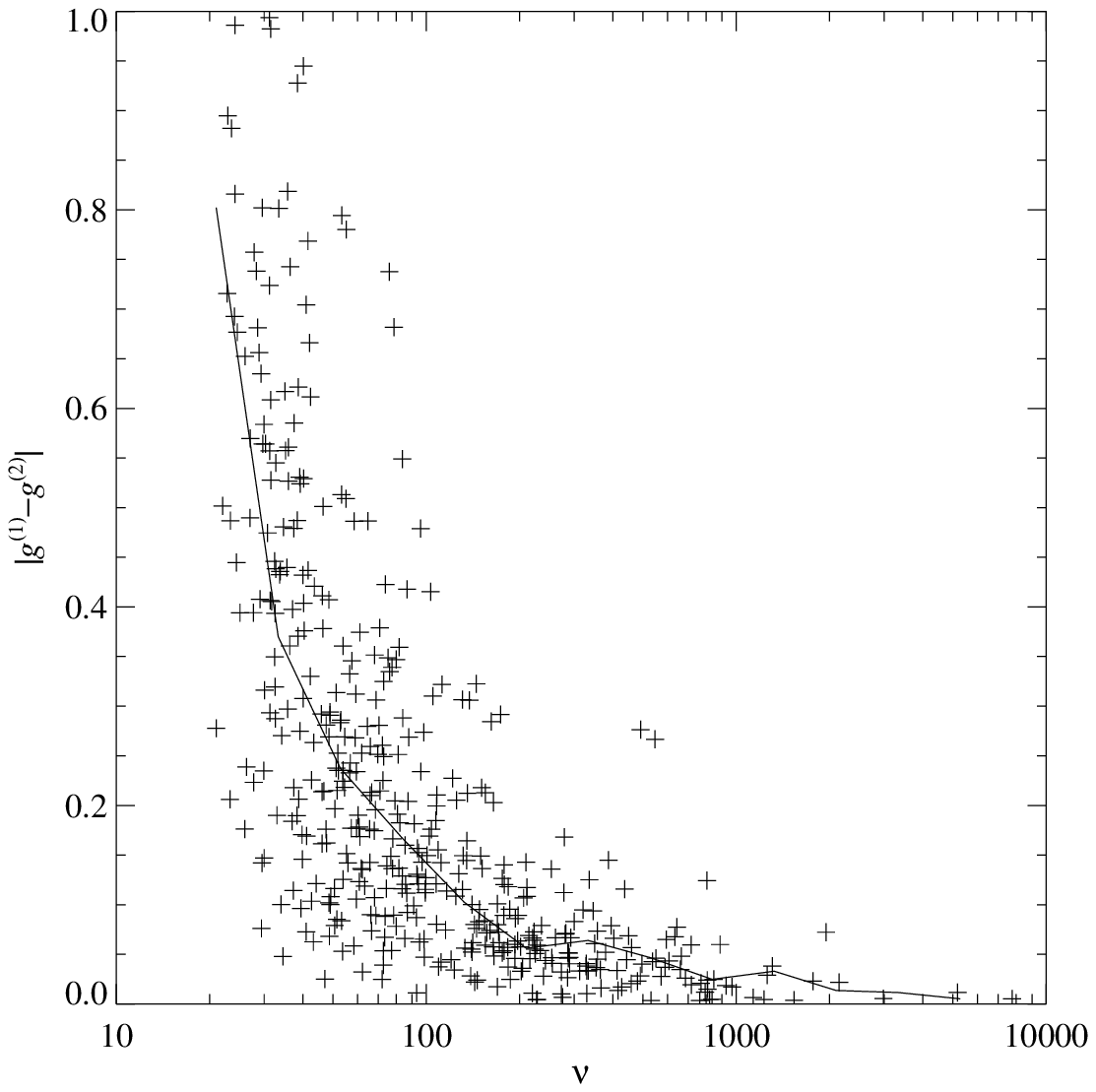}
  \caption{Photometric errors on the shear measurements on galaxies
    observed in two adjacent pointings.  Left: differences in the
    shear real and imaginary parts; note that the errors are
    isotropic.  Right: the error on the shear measurements as a
    function of the detection significance $\nu$ of the galaxy.  The
    solid line shows the average error observed for each value of
    $\nu$.}
  \label{fig:4}
\end{figure*}

Since source ellipticities are expected to vanish on average, we
obtained an estimate of the complex shear $g$ acting on each galaxy by
taking $\chi^0 = 0$ and inverting Eq.~\eqref{eq:2}.  In order to
estimate the error on the shear measurement of each galaxy, we
calculated the intrinsic dispersion on the ellipticities for similar
galaxies.  This was done by calculating the dispersion on the measured
shear $\sum |g|^2 / N$ for objects having similar sizes and
magnitudes.  This dispersion comprises the ellipticity uncertainty due
to the photometric error and the intrinsic galaxy ellipticity (the
average departure of galaxies from a circular shape), which
effectively acts as a source of error for the shear measurements.  The
photometric error can be conveniently estimated using multiple (two to
four) measurements in the overlapping regions by evaluating the
differences in the ellipticity of the same galaxy.  As an example, in
Fig.~\ref{fig:3} we show the measured shears of galaxies identified on
two adjacent pointings.  Interestingly, this plot shows that, for the
faint galaxies used in our weak lensing analysis, the photometric
error is a significant fraction of the total (photometric and
intrinsic) ellipticity error.  For example, for the $i$ band shear
estimates we measure an average photometric error
$\mbox{Err}_\mathrm{phot}(g) = 0.148$ and an intrinsic scatter of
ellipticity $\mbox{Err}_\mathrm{intr}(g) = 0.286$ (since the two
errors are independent, the total scatter in shear is about
$\bigl\langle | g |^2 \bigr\rangle^{1/2} = 0.322$).
Figure~\ref{fig:4} shows the photometric errors $\Delta g$ on the
complex plane, and the distribution of photometric errors as a
function of the detection significance.

Finally, we merged the four catalogs and replaced multiple entries in
the overlapping regions with single ellipticities obtained from a
weighted average of the measured shears, and errors evaluated by
properly separating the statistical (photometric) and systematic
(intrinsic galaxy ellipticity) errors.

\subsection{Selection of background galaxies}
\label{sec:backgr-galaxy-select}

\begin{figure*}[!t]
  \begin{center}
    \includegraphics[bb=22 213 573 629, width=\hsize,
    keepaspectratio]{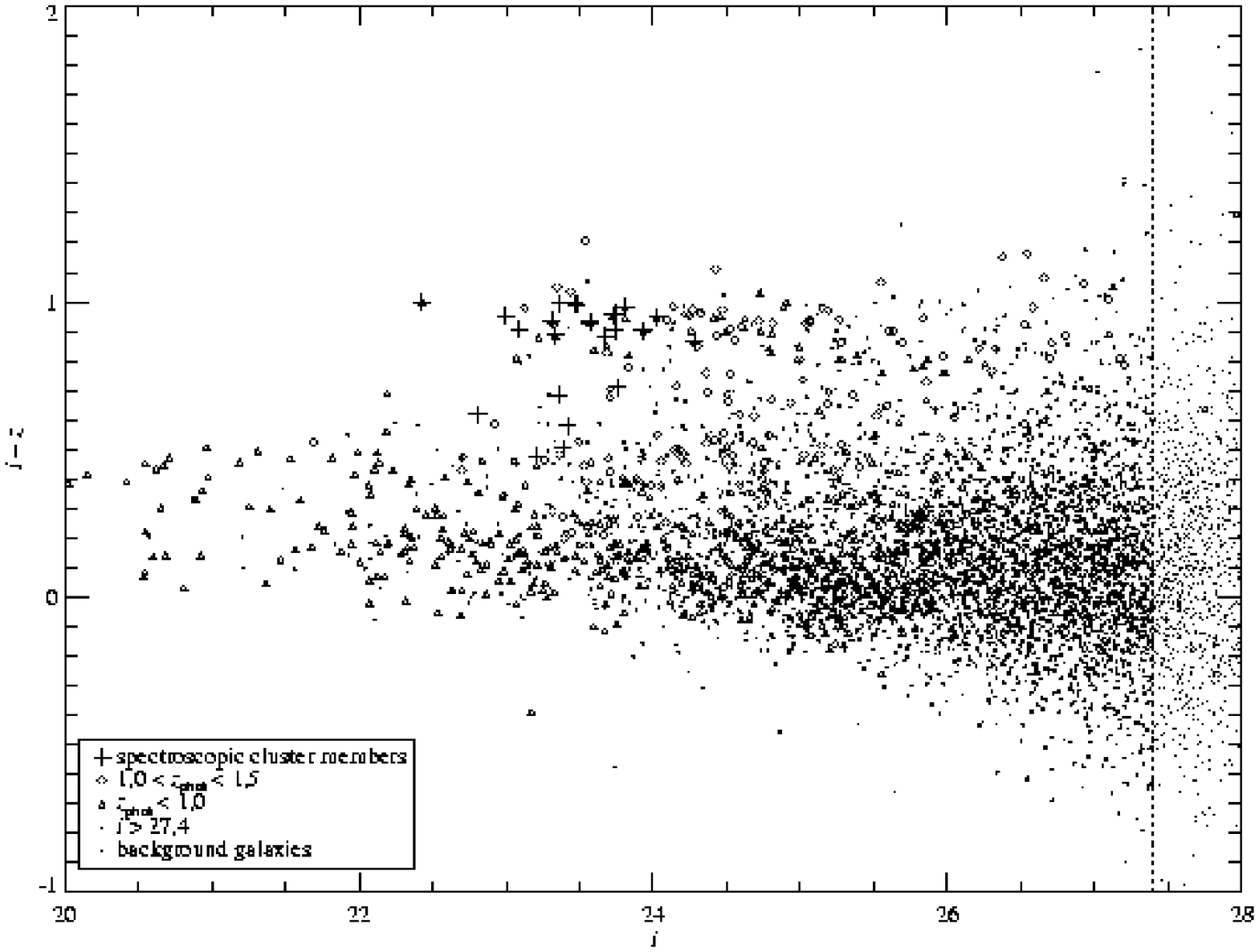}
    \caption{The color-magnitude diagram for the galaxies in \cluster\
      field.  Objects effectively used for the lensing analysis are
      marked as ``background galaxies.''}
    \label{fig:5}
  \end{center}
\end{figure*}

We classified galaxies as background or foreground with respect to the
cluster using ground-based photometric redshifts as described in
\citet{2004astro.ph..4474T}.  In particular, we matched the two
catalogs obtained from the analysis described above with a catalog
containing photometric and spectroscopic redshifts for $\sim 1\,$200
objects (see Fig.~\ref{fig:5}).  We then selected as fiducial
background objects all galaxies on the photometric redshift catalog
with $z_\mathrm{phot} > 1.5$; other galaxies too faint to be included
in the photometric redshift catalog were taken to be background and
included in the final catalog.  Finally, we conservatively discarded
for the $i$-band (respectively, $z$-band) all galaxies with magnitude
$i > 27.4$ ($z > 26.5$) because, in most cases, these objects were too
faint to provide reliable shear measurements.  These two final
catalogs were visually inspected and used for the weak lensing
analysis discussed below.  The $i$ band and $z$ band catalogs contain,
respectively, $3\,980$ and $2\,370$ galaxies, corresponding to about
$120$ and $70 \mbox{ galaxies arcmin}^{-2}$.  We studied the dependence
of the weak lensing signal for different magnitude and photometric
redshift cuts, and verified that its strength was maximized for the
set of parameters chosen above to select the background galaxy
catalog.  Interestingly, we were still able to measure a lensing
signal when all galaxies were taken to be background and included in
the input catalog, although in this case the lensing signal is
depressed because of the dilution by foreground galaxies.
Figure~\ref{fig:5} shows the different galaxy subsamples in a
color-magnitude plot, with $i - z$ derived from ACS isophotal
magnitudes.

\section{Results}
\label{sec:results}

\subsection{Mass maps}
\label{sec:mass-maps}

\begin{figure*}[!t]
  \begin{center}
    \includegraphics[bb=76 211 519 630, width=0.48\hsize,
    keepaspectratio]{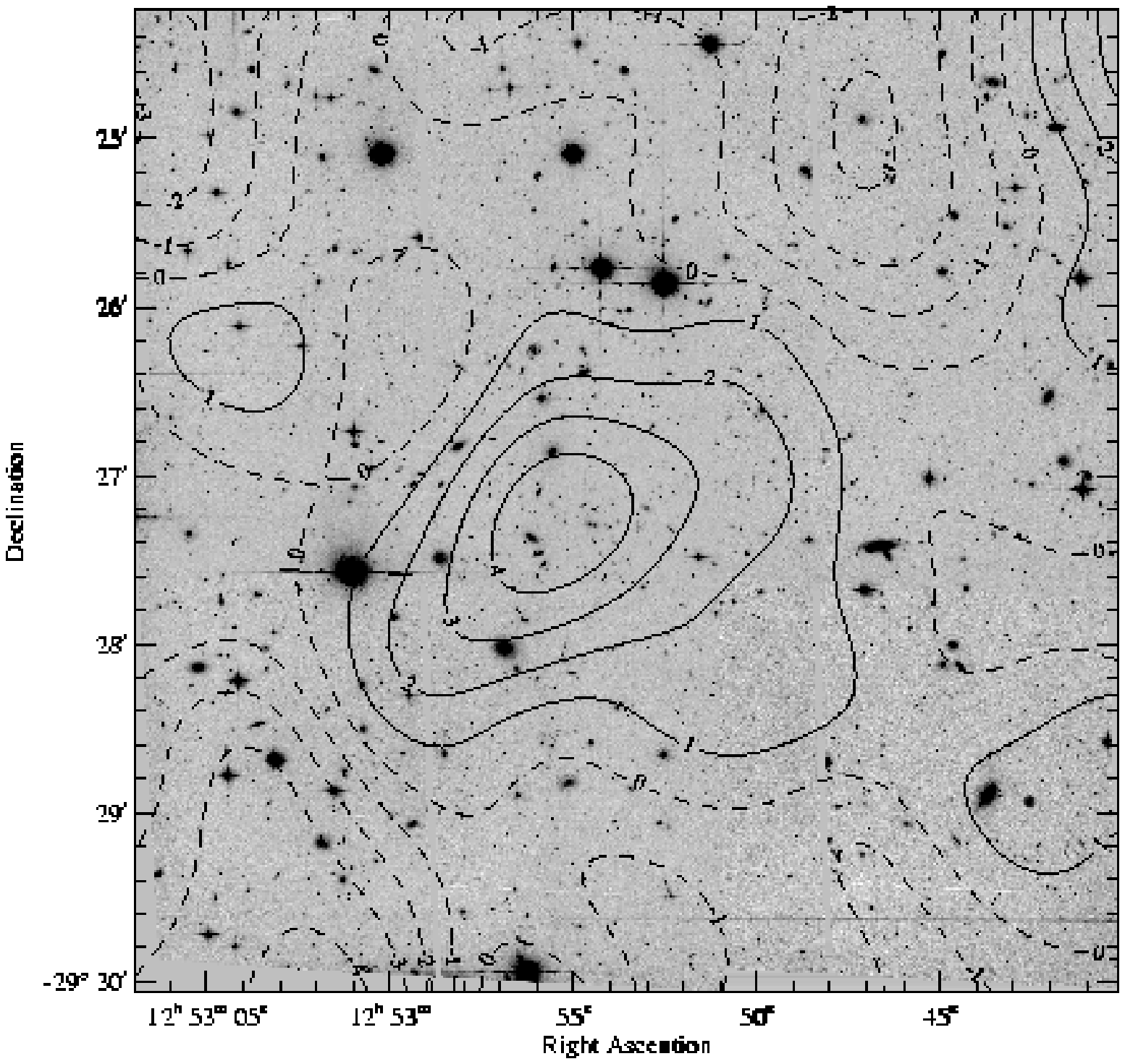}
    \hfill
    \includegraphics[bb=76 211 519 630, width=0.48\hsize,
    keepaspectratio]{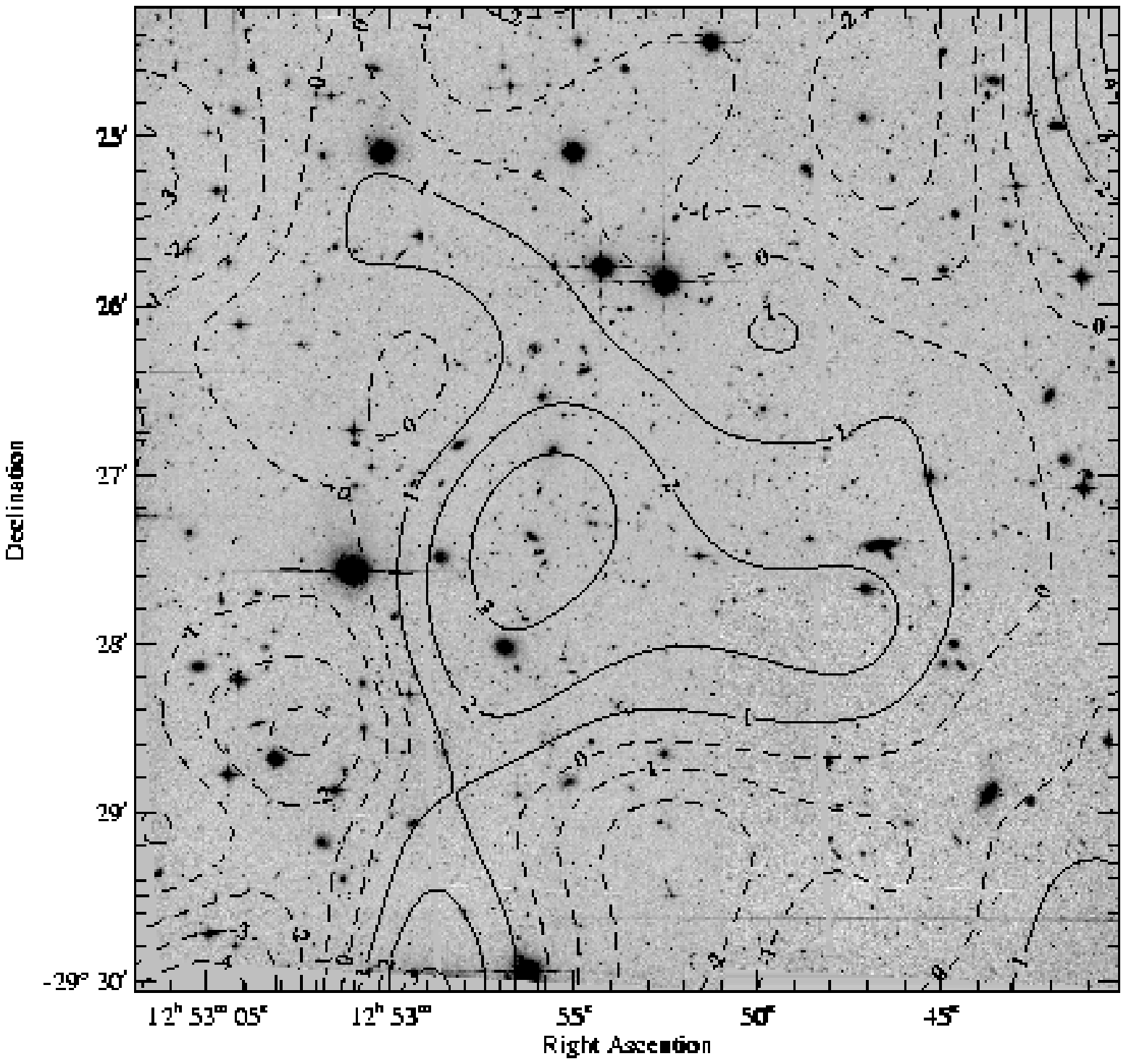}
    \caption{Contours of the dimensionless mass maps in the $i$ (left)
      and $z$ (right) bands, overlaid onto the $i$-band ACS image.
      Contours levels are spaced by $0.019$ for the $i$ map and
      $0.025$ for the $z$ map.  These values correspond to the
      estimated median errors of the two maps over the field (see
      Fig.~\ref{fig:7}).  Solid (respectively, dashed) contours
      represent overdensities (underdensities) with respect to the
      average over the whole field.  The mass sheet degeneracy is not
      removed at this stage, and the mass maps are normalized so that
      the integrals of them over the whole field vanish.}
    \label{fig:6}
  \end{center}
\end{figure*}

\begin{figure*}[!t]
  \begin{center}
    \includegraphics[bb=94 216 500 618, width=0.48\hsize,
    keepaspectratio]{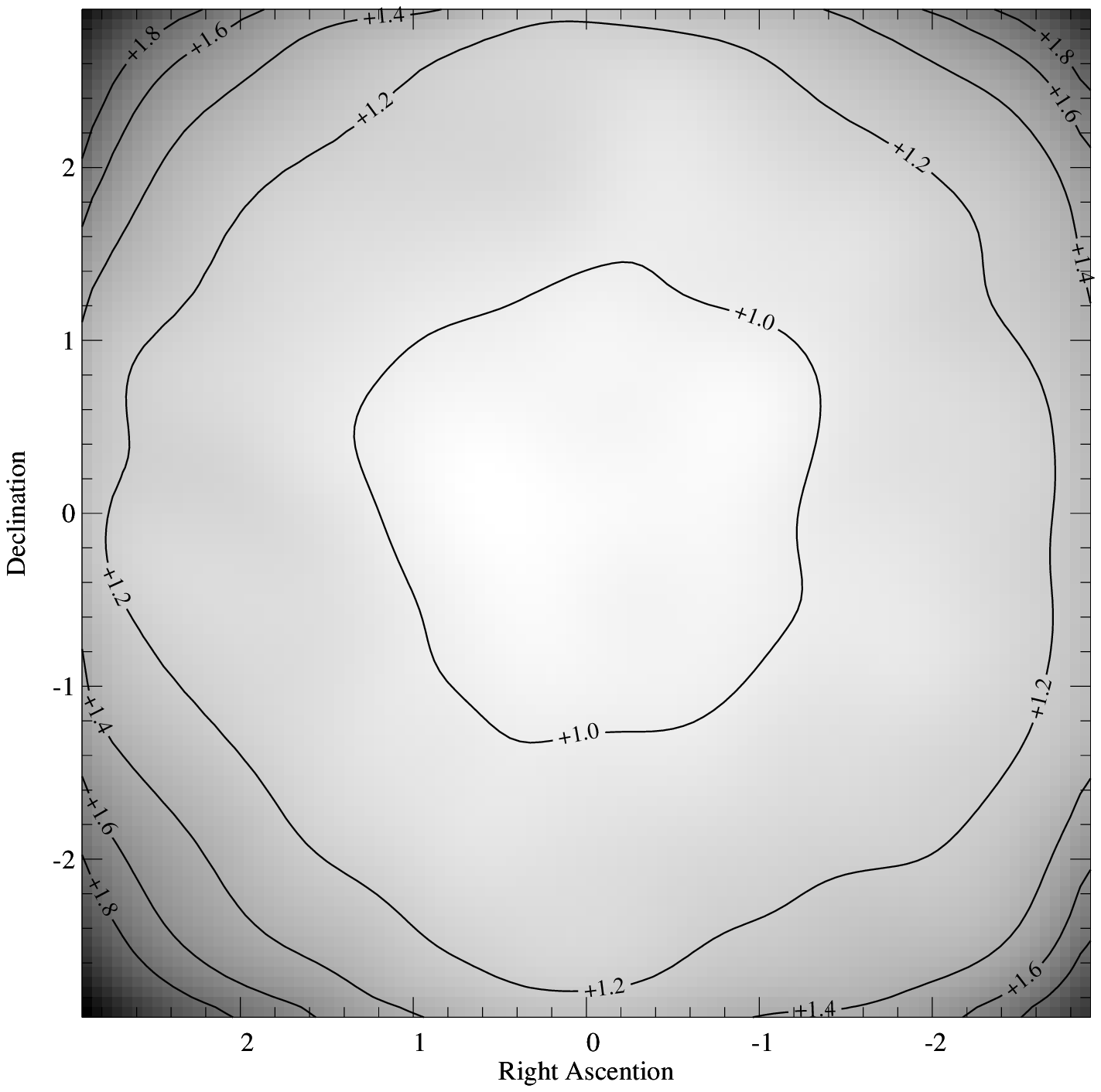}
    \hfill
    \includegraphics[bb=94 216 500 618, width=0.48\hsize,
    keepaspectratio]{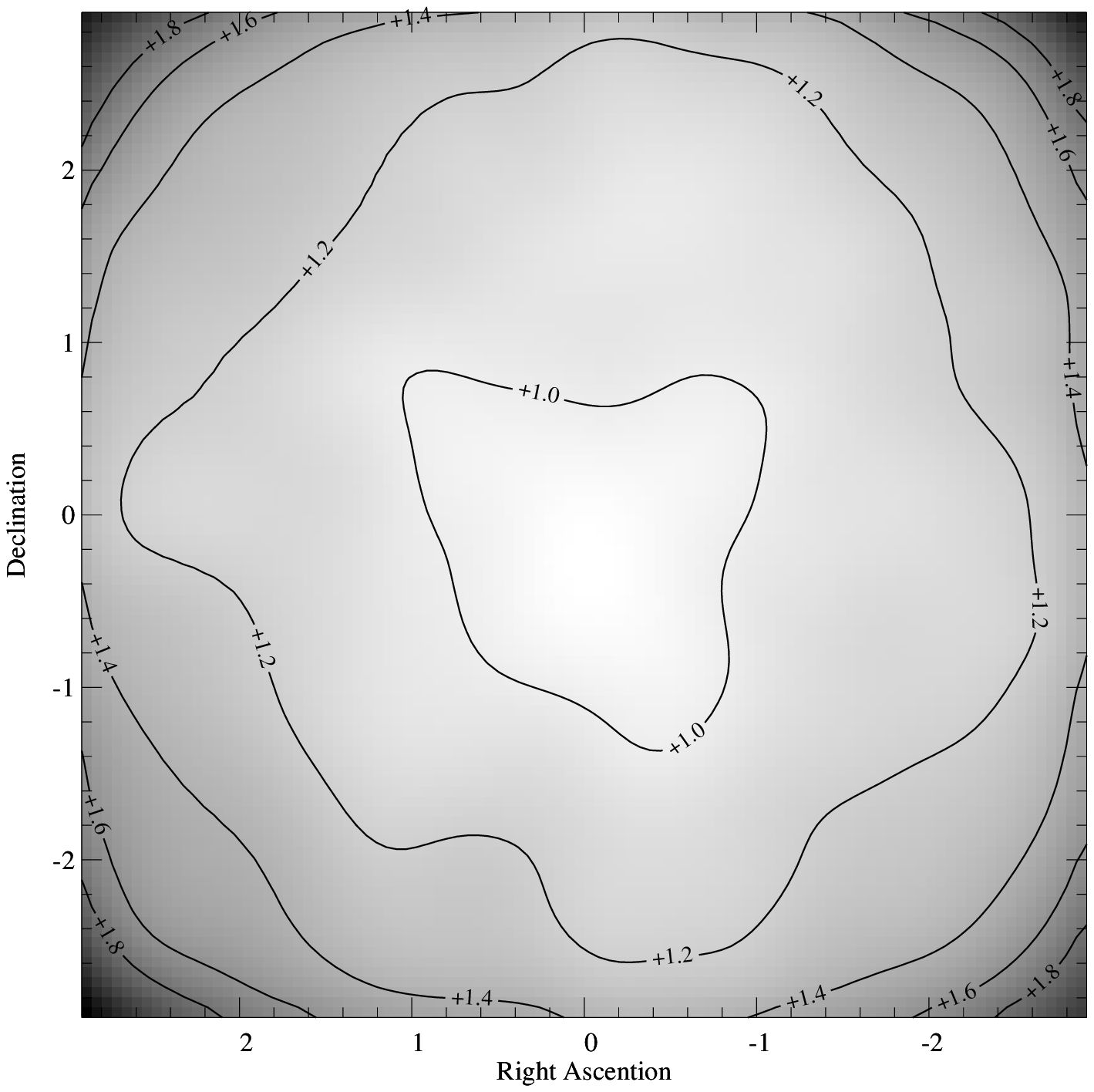}
    \caption{The expected local error on the $i$ (left) and $z$
      (right) mass maps shown in Fig.~\ref{fig:6}.  Contours are
      labeled in units of median value over the whole field.  As
      expected, the error at the center of the field is below the
      average, while it increases significantly (by a factor
      $1.6$--$1.8$) at the corners.  The comparison of these maps with
      the structures observed in Fig.~\ref{fig:6} allows us to assess
      the detection significance of substructures.}
    \label{fig:7}
  \end{center}
\end{figure*}

\begin{figure*}[!t]
  \begin{center}
    \includegraphics[bb=76 211 519 630, width=\hsize, 
    keepaspectratio]{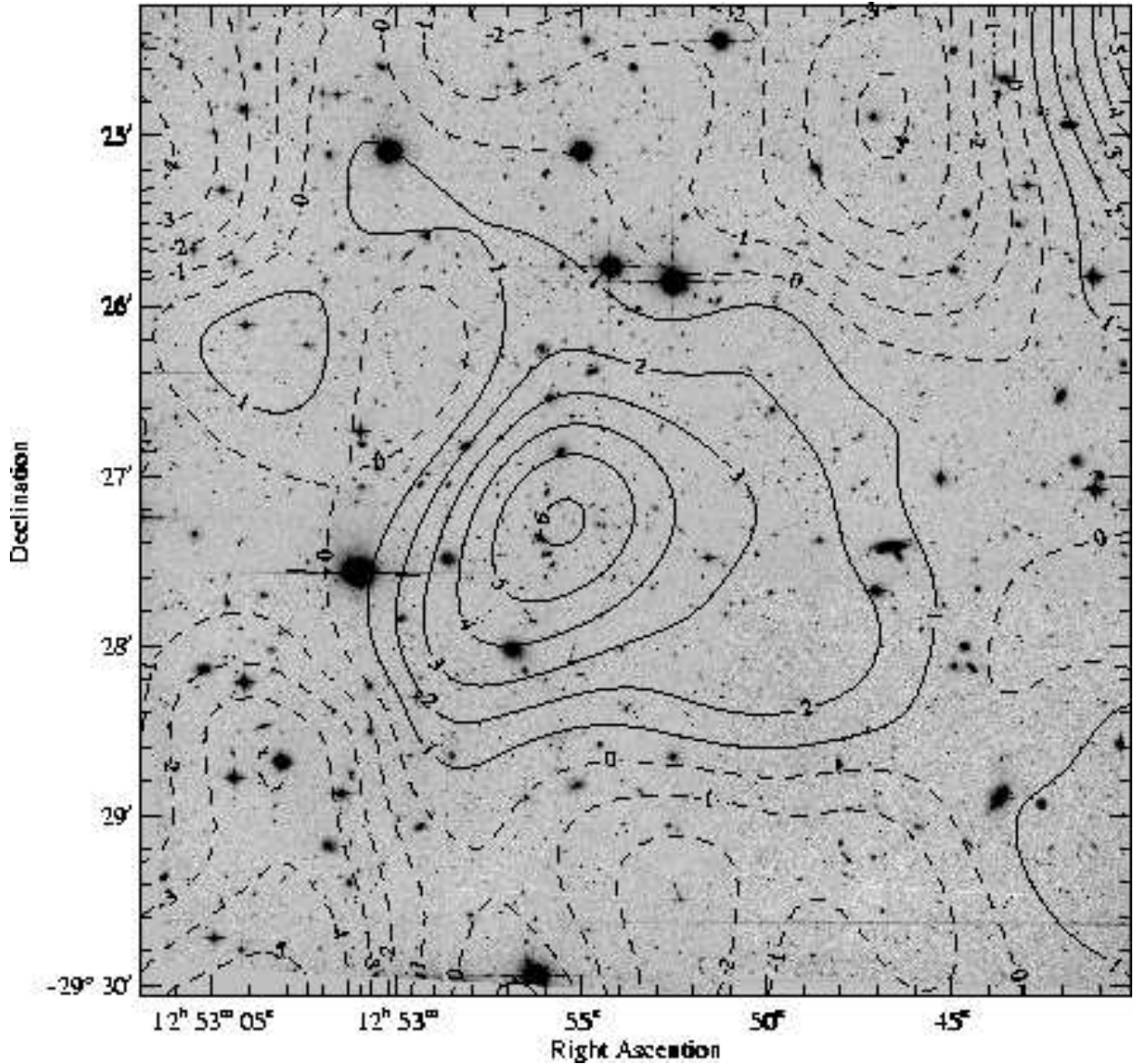}
    \caption{The optimally combined mass map of \Cluster\ using both
      $i$ and $z$ ellipticities measurements.  Contours are spaced by
      $0.015$, corresponding to the estimated median error of the map
      over the field.}
    \label{fig:8}
  \end{center}
\end{figure*}

\begin{figure*}[!t]
  \begin{center}
    \includegraphics[bb=76 211 519 630, width=\hsize, 
    keepaspectratio]{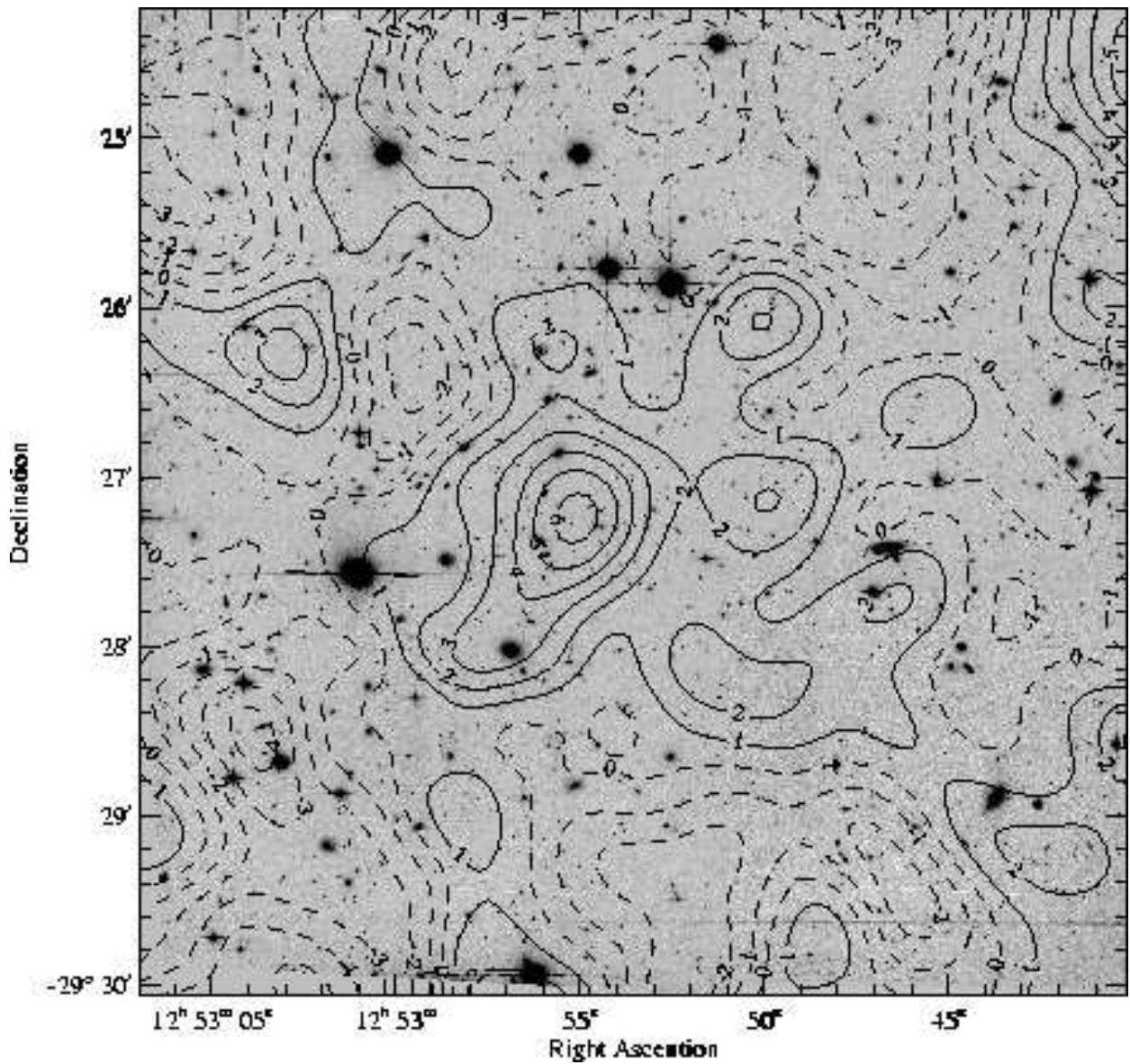}
    \caption{Higher resolution mass map of Fig.~\ref{fig:8},
      obtained with a smoothing scale of $\sigma_\mathrm{W} = 15
      \mbox{ arcsec}$.}
    \label{fig:9}
  \end{center}
\end{figure*}

\begin{figure}[!t]
  \begin{center}
    \includegraphics[bb=94 216 498 617, width=\hsize,
    keepaspectratio]{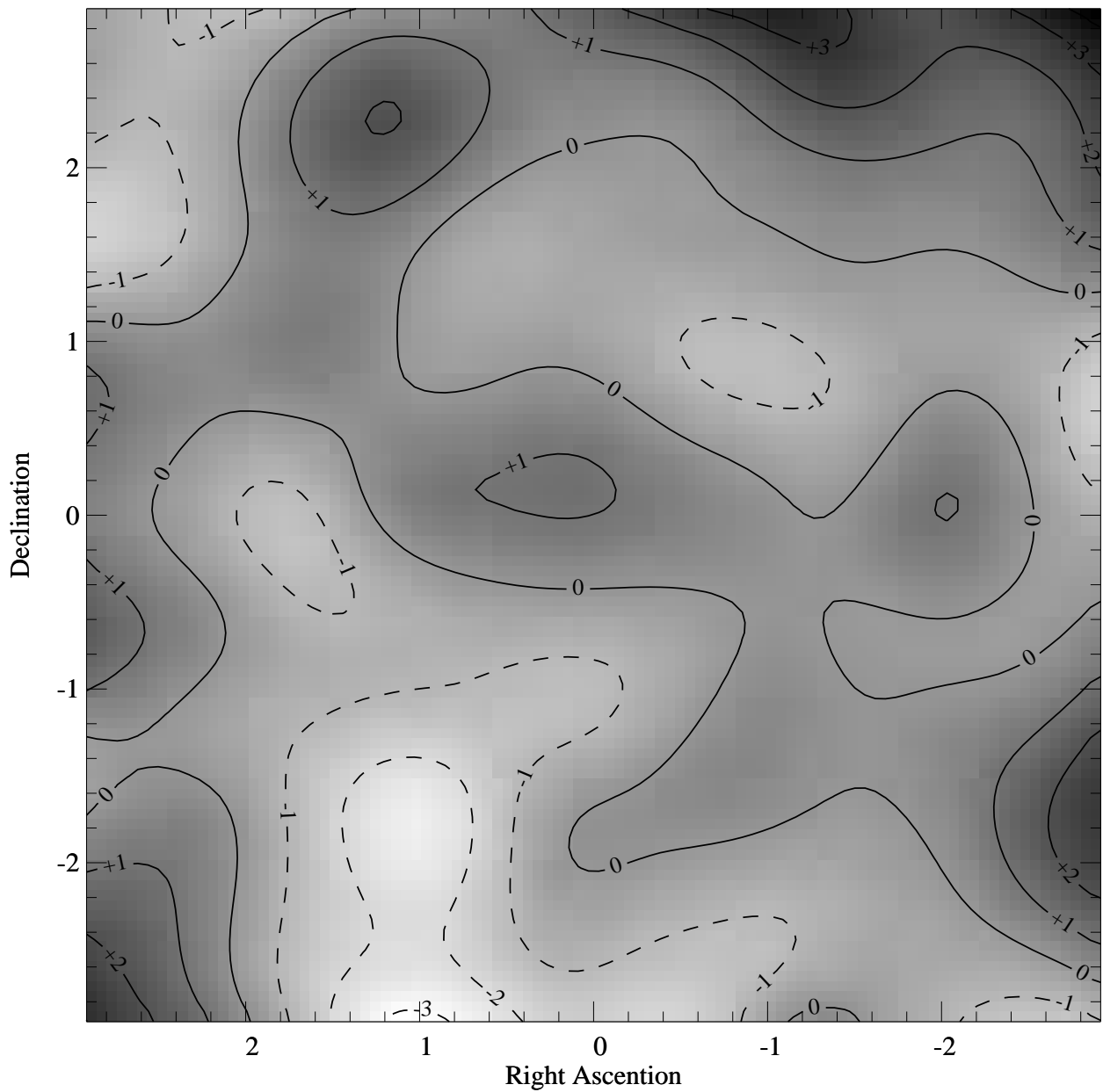}
    \caption{An example of realization of a bootstrap shear map in the
      $i$ band.  Note that the reconstructed mass distribution is
      consistent with a zero at 1-$\sigma$ level.}
    \label{fig:10}
  \end{center}
\end{figure}

\begin{figure*}[!t]
  \begin{center}
    \includegraphics[bb=94 216 499 617, width=0.48\hsize,
    keepaspectratio]{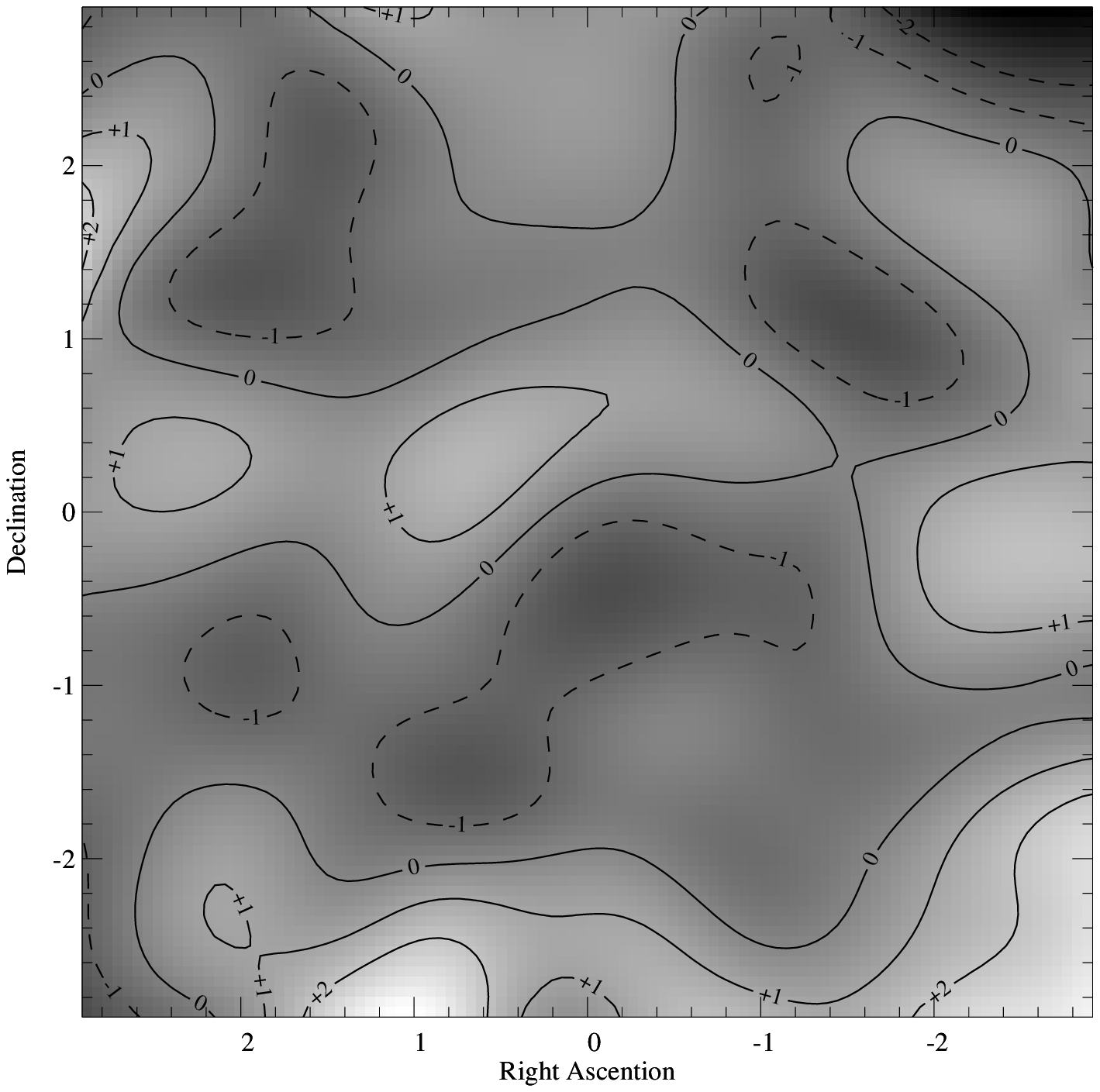}
    \hfill
    \includegraphics[bb=94 216 499 617, width=0.48\hsize,
    keepaspectratio]{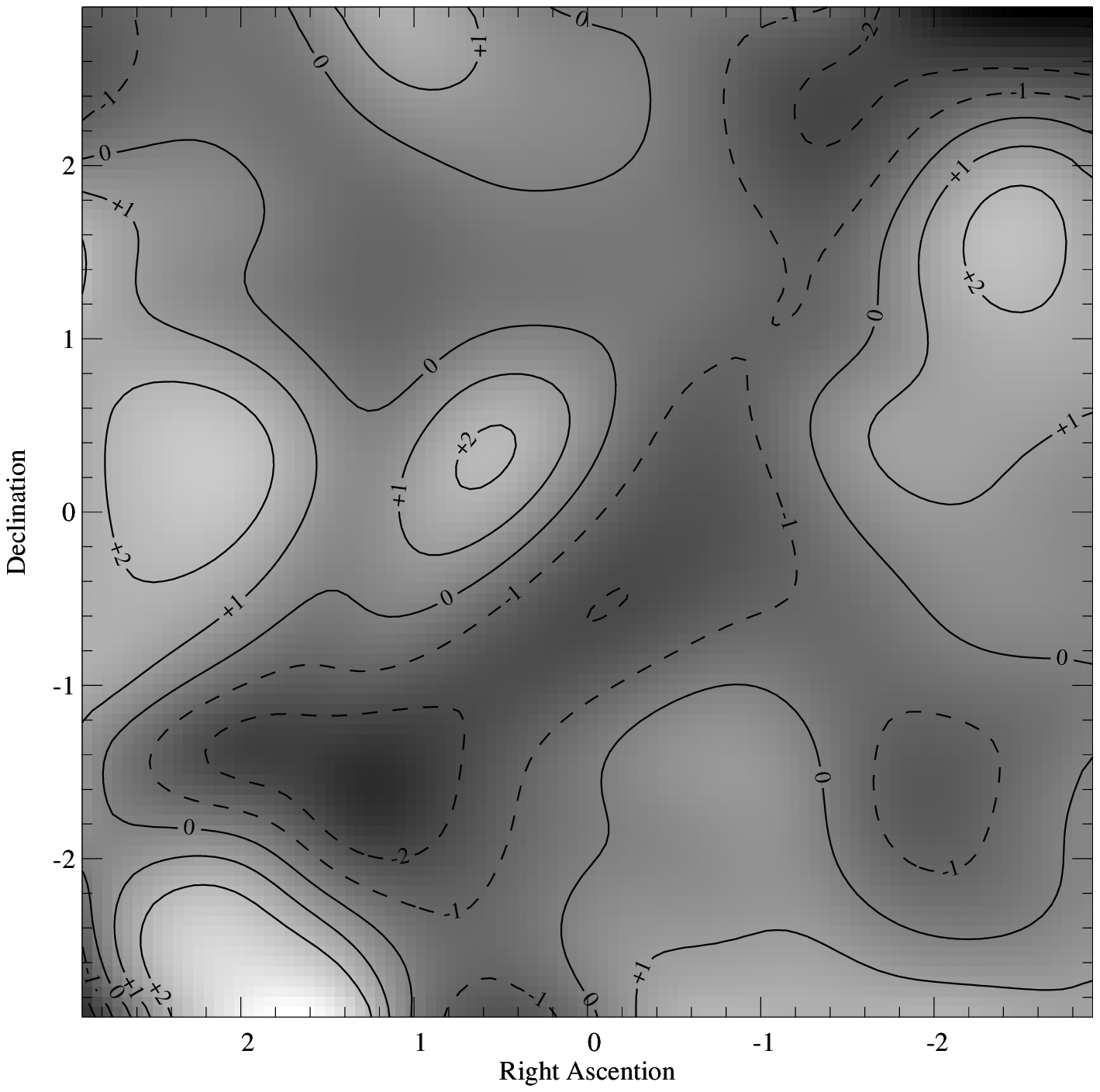}
    \caption{The mass distribution obtained by applying the
      transformation \eqref{eq:4} to all galaxy ellipticities in the
      $i$ (left) and $z$ (right) band.  Similarly to the other
      figures, the contours are spaced by 1-$\sigma$.  Since the
      lensing shear is curl-free, the reconstructed mass map is
      expected to vanish, which is what we observe (see text).}
    \label{fig:11}
  \end{center}
\end{figure*}

In order to obtain the projected mass distribution of the cluster, we
smoothed the galaxy shear estimates into a continuous shear field; the
smoothing was performed using a Gaussian kernel characterized by
dispersion $\sigma_\mathrm{W} = 25 \mbox{ arcsec}$.  We inverted the
shear field into the convergence using the optimal finite-field
inversion algorithm \citep[see][]{1996A&A...305..383S,
  1998A&A...335....1L}, with the implementation described by
\citet{1999A&A...348...38L}.  The results are shown in
Fig.~\ref{fig:6} for both bands.  

We also estimated the local error on these maps using two different
methods, (i) an analytical estimate based on the measured errors on
the measured shears for each galaxies
\citep[see][]{1998A&A...335....1L} and (ii) a ``bootstrap estimate''
obtained by assigning galaxy ellipticities drawn from the original
catalog to randomly selected galaxy positions \citep[e.g.][]{Efron82}.
These two estimates are in excellent agreement, and give robust
statistical significance for the weak lensing detection of \cluster\ 
described below.  Thus, we could verify that we obtained a
$5$-$\sigma$ lensing signal on the $i$ band and a $3$-$\sigma$
detection on the shallower $z$ band.  The predicted analytical errors
in the $i$ and $z$ maps are shown in Fig.~\ref{fig:7} (note that
errors are expected to be correlated on the scale $\sigma_\mathrm{W}$
of the Gaussian smoothing applied on the field).  As expected, the
error is smaller at the center of the field, which is deeper and free
from boundary effects.  Using these error estimates, we could also
optimally combine the $i$ and $z$ mass reconstructions into a single
field, shown in Fig.~\ref{fig:8}.  The detection significance in this
combined map is $6$-$\sigma$.  We also evaluated the weak lensing mass
map using a Gaussian kernel with smaller dispersion,
$\sigma_\mathrm{W} = 15 \mbox{ arcsec}$.  The result, shown in
Fig.~\ref{fig:9}, shows evidence of substructure (note, in
particular, the elongation North-South of the main clump, and the
presence of several smaller clumps to the West of the cluster core).

In order to check the reliability of our weak lensing detection, we
performed several tests.  Figure~\ref{fig:10} shows an example of a
single realization of mass map from a bootstrapped catalog (generated
using the $i$ band catalog); note that the observed convergence is
consistent with a vanishing field.  A more stringent test for
systematic effects was obtained by rotating all galaxy ellipticities
by $45^\circ$, i.e. by operating the transformation
\begin{equation}
  \label{eq:4}
  (\chi_1, \chi_2) \mapsto (\chi_2, -\chi_1) \; .
\end{equation}
and by evaluating the resulting mass distribution $\kappa_\times$.
Because of the properties of the lens mapping, the field
$\kappa_\times$ should vanish; moreover, the noise properties of field
are very similar to the ones of $\kappa$.  In this respect we note
that, in principle, the noise on the $\kappa_\times$ map is slightly
larger, because of the increase on scatter of the lensed ellipticities
introduced by the cluster; however, in practice, this effect is
expected to be very small for a weak lens such as \cluster.  The
$\kappa_\times$ map is shown in Figs.~\ref{fig:11} for the $i$ and $z$
bands.  Clearly the $i$ band does not show any significant peak, while
the $z$ band has a few $2$-$\sigma$ peaks, still consistent with a
vanishing field.  Given the size of the smoothing performed compared
with the size of the field, we have approximately $10 \times 10$
independent measurements on the whole area; the $2$-$\sigma$ peaks, in
total, occupy about $10\%$ of the field, which is what we would expect
from simple statistical arguments.

To further test the statistical significance of our result, we
considered the integral 
\begin{equation}
  \label{eq:5}
  I_\times = \int_\Omega \bigl[ \kappa_\times(\vec\theta)
  \bigr]^2 \, \diff^2 \theta \; , 
\end{equation}
over the field $\Omega$ (excluding regions close to the edges of the
field in order to avoid boundary effects).  The quantity $I_\times$
defined above is clearly always positive, and measures the departure
of the reconstructed field from a null field.  Given its analytical
expression, this quantity is expected to behave as a $\chi^2$-like
distribution.  In absence of any systematic effect, $\kappa_\times$
should be consistent with a vanishing field, and thus to have
departures from zero consistently with the bootstrapped maps.  In
other words, $\kappa_\times$ should be statistical indistinguishable
from a bootstrapped map.  In order to test this, we compared the value
of $I_\times$ obtained from $\kappa_\times$ with the analogous
quantity obtained from $1\,000$ realizations of bootstrapped catalogs.
Reassuringly, our analysis shows that in a significant fraction of
cases ($41\%$ for the $z$ band and $93\%$ for the $i$ band) we
measured values for $I_\times$ from the bootstrapped maps larger than
those obtained from our data.  This confirms that both $i$ and $z$
band cross convergences are consistent with a zero.

We also evaluated the integral \eqref{eq:5} using $\kappa$ instead of
$\kappa_\times$, and compared the values obtained with the results of
the bootstrapping.  In the $i$ band, we had in one single case (out of
$1\,000$) a bootstrapping map with a value exceeding the one observed;
in the $z$ band we had 20 cases out of $1\,000$; finally, comparing
the combined $i$ + $z$ map and the similar combined bootstrapped maps,
we never saw a signal comparable with what we observed. 

\begin{figure*}[!t]
  \begin{center}
    \includegraphics[width=\hsize,keepaspectratio]{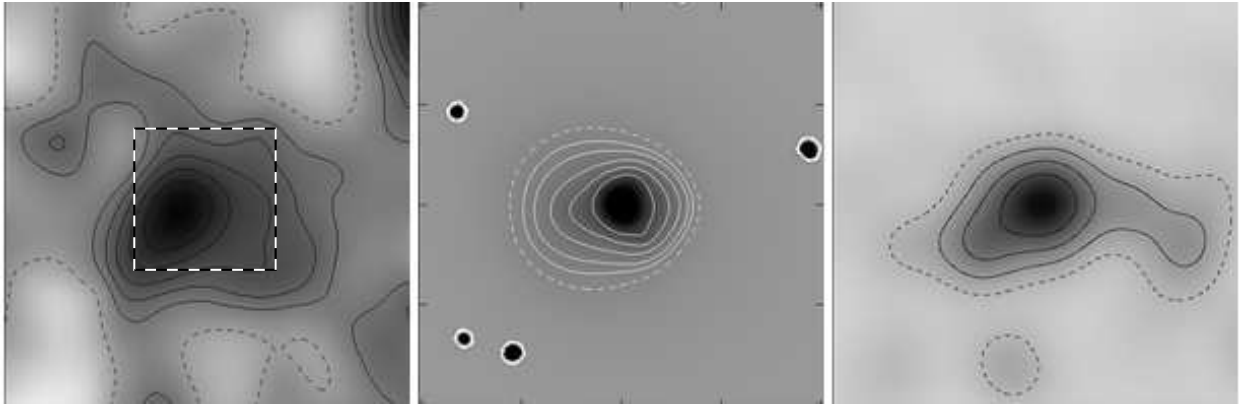}%
    \begin{pspicture}(\hsize,0)(\hsize,0)%
      \psset{unit=0.33\hsize}
      \psframe(0.325,0.325)(0.675,0.675)
      \psframe[linestyle=dashed,dash=4pt 4pt,linecolor=white](0.325,0.325)(0.675,0.675)
    \end{pspicture}%
    \caption{From left to right: the weak lensing mass distribution of
      Fig.~\ref{fig:8}; the X-ray emitting gas contours, obtained with
      an adaptive smoothing of the \textit{Chandra\/} observation in
      the $1$--$2 \mbox{ keV}$; the $K$ band luminosity weighted light
      distribution \citep{2004astro.ph..4474T}.  All panels are
      centered on the optical center of the cluster; the lensing map
      is $5.7 \times 5.7 \mbox{ arcmin}^2$ wide, while the other
      panels have a side length of $2 \mbox{ arcmin}$ (or 1 Mpc),
      corresponding to the dashed square in the first panel.}
    \label{fig:12}
  \end{center}
\end{figure*}

The overall lensing maps, and in particular Figs.~\ref{fig:8} and
\ref{fig:9}, show a clear elongation of the mass distribution in
direction East-West.  Although it is generally difficult to evaluate
the significance of substructure detections with weak lensing
\citep[see, e.g.][]{2002MNRAS.335.1037M}, we note \textit{a
  posteriori\/} that this elongation is consistent with what inferred
from the X-ray and optical contours (see Fig.~\ref{fig:12} and
discussion in Sect.~\ref{sec:param-mass-estim}).  As discussed in
\citet{2004AJ....127..230R}, the X-ray surface brightness distribution
suggests that we might be observing \cluster\ in a post-merging phase
along the East-West direction. Interestingly, both the shear map and
the azimuthally averaged mass map (see Fig.~\ref{fig:23} below) extend
well beyond the detected X-ray emission, which is affected by severe
surface brightness dimming at these redshifts.

In order to convert the dimensionless convergence map into a real mass
density field we needed to (i) estimate the redshift distribution of
background galaxies and (ii) remove the so-called mass-sheet
degeneracy.  We will consider these issues in the next two
subsections.

\subsection{The background redshift distribution}
\label{sec:backgr-redsh-distr}

As discussed in Sect.~\ref{sec:basic-relations}, the measured
(reduced) shear is a function of the redshifts of the background
galaxies.  Since photometric redshifts are available only for the
brightest galaxies (the ones detected on the ground-based images), we
estimated the background galaxy redshift distribution $p(z)$ by
resampling the photometric redshift catalogs of the Hubble Deep
Fields.  We used the catalogs provided by \citet{1999ApJ...513...34F}
and \citet{2002ApJ...570..492L}, and also a newly generated catalog
based on the Bayesian photometric redshift estimation
\citep{2000ApJ...536..571B}.  Specifically, we matched these catalogs
(obtained from HST/WFPC2 observations) with ACS GTO observations in
the $F775W$ and $F860LP$.  Thus, we obtained photometry for a sample
of approximately $1\,000$ objects with photometric redshifts in the
same filters as \cluster\ observations.  We then divided the HDF
catalogs in magnitude bins and derived the $p(z)$ of galaxies in our
field by associating the HDF photometric redshifts in the
corresponding bin.  Such a method to estimate $p(z)$ is generally
affected by cosmic variance because of the small area covered by the
HDFs.  We estimate this uncertainty below by comparing the $p(z)$
derived from the HDF-North and South.

\begin{figure}[!t]
  \begin{center}
    \includegraphics[bb=144 316 454 518, width=\hsize, 
    keepaspectratio]{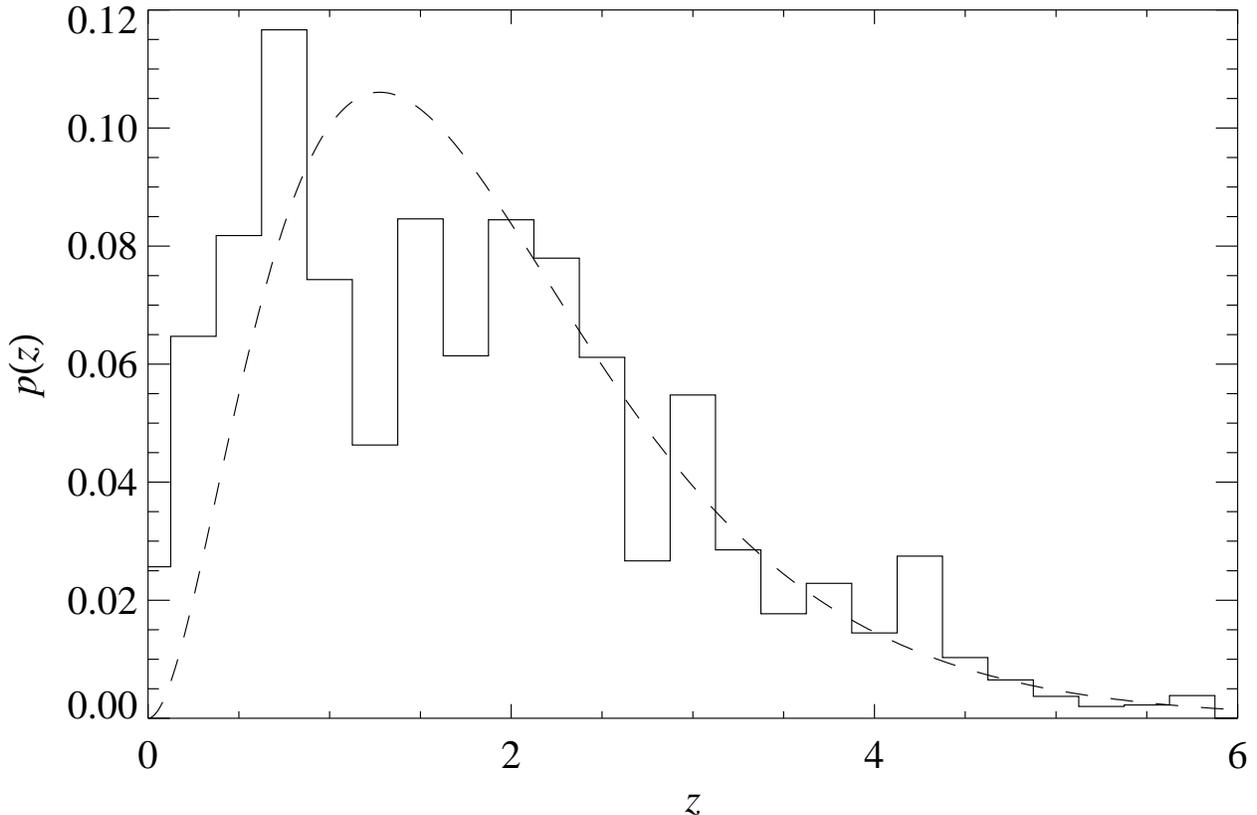}
    \caption{The histogram shows the redshift distribution of the
      galaxy catalog included in the weak lensing analysis, as derived
      by resampling the Hubble Deep Field North photometric redshift
      catalog (analyzed using the method by
      \citealp{2000ApJ...536..571B}; see text); the dashed smooth line
      shows the best fit with simple model for $p(z)$ of the form
      \eqref{eq:7} with $z_0 = 1.28$.  Note that the peak of the
      distribution (mode) is at $z = z_0$ and the average is $\langle
      z \rangle = (3 / 2) z_0$.}
    \label{fig:13}
  \end{center}
\end{figure}



The application of this method to the HDF North led to the redshift
distribution shown in Fig.~\ref{fig:13} and to an \textit{effective\/}
background source redshift $z_\mathrm{eff}$. This quantity is defined
as the redshift at which all galaxies would need to be in order to
produce the same (average) lensing signal as the galaxies distributed
as $p(z)$:
\begin{equation}
  \label{eq:6}
  \bigl[ \Sigma_\mathrm{c}(z_\mathrm{eff}) \bigr]^{-1} = \int p_z(z) \bigl[
  \Sigma_\mathrm{c}(z) \bigr]^{-1} \, \diff z \; .
\end{equation}
Using the three HDFs catalogs, we obtained respectively
$z_\mathrm{eff} = 1.61$ (HDF-N; \citealp{1999ApJ...513...34F}),
$z_\mathrm{eff} = 1.72$ (HDF-N Bayesian estimation;
\citealp{2000ApJ...536..571B}), and $z_\mathrm{eff} = 1.81$ (HDF-S;
\citealp{2002ApJ...570..492L}).  We used $z_\mathrm{eff} = 1.72$ as
best estimate, and estimated the error on the effective redshift to be
approximately $0.1$.  We can approximate the galaxy redshift
distribution with a simple function of the form \citep[see,
e.g.][]{1999A&A...342..337L}
\begin{equation}
  \label{eq:7}
  p(z) = \frac{4 z^2}{z_0^3} \e^{-2 z / z_0} \; ,
\end{equation}
and obtained a best fit value $z_0 = 1.28$ (see Fig.~\ref{fig:13}).
Although we did not directly use this parametrization in our study,
but rather the resampling technique described above, the use of $p(z)$
in the form of Eq.~\eqref{eq:7} better elucidates the estimated depth
of our observations.  Note that $z_\mathrm{eff}$, which controls
directly the scaling between the dimensionless mass distribution
$\kappa(\vec\theta)$ and the dimensional one $\Sigma(\vec\theta)$,
strongly depends on the lens redshift $z_\mathrm{d} = 1.24$; the
parameter $z_0$, instead, only characterizes the background galaxy
redshift distribution, and thus can be better used to compare weak
lensing studies performed on different clusters.  Note that an
uncertainty on $z_\mathrm{eff}$ of $0.1$ corresponds to a $\sim 15\%$
uncertainty in final mass estimate.


\subsection{Parametric mass estimates}
\label{sec:param-mass-estim}

In order to break the mass-sheet degeneracy, which is an intrinsic
limitation of parameter-free mass reconstructions, and to better
investigate the mass profile, we fitted our data with parametric mass
models.

We used both the $i$ and $z$ band catalogs, and fitted directly the
observed ellipticities with the shear field predicted by parametric
mass models. We stress that this allows us to proceed without
smoothing, which would be otherwise needed if we were fitting the
reconstructed mass distributions (as discussed in
Sect.~\ref{sec:mass-maps}) or the shear profiles (see below
Sect.~\ref{sec:shear-profile}).  In particular, we used the chi-square
function
\begin{equation}
  \label{eq:8}
  \chi^2 = \sum_{n=1}^N \frac{\bigl\lvert g_n - g(\vec\theta_n)
    \bigr\rvert^2} {\sigma^2_n} \; ,
\end{equation}
where $g_n$ is the measured shear for the $n$-th galaxy, $\sigma_n$ is the
associated measurement error (estimated as discussed in
Sect.~\ref{sec:ellipt-meas}), $\vec\theta_n$ is the galaxy location on
the sky, and $g(\vec\theta)$ is the shear predicted by the model chosen
(which depends on the model parameters).

Although the mass map shows a slight elongation in direction
East-West, we decided to use axisymmetric mass models which are easier
to interpret and to compare with independent mass estimates (see
below).  Thus we used four different models:
\begin{description}
\item[NISwl] A non-singular isothermal sphere centered at the maximum
  of the weak lensing mass map ($\mbox{R.A.: }12^\mathrm{h}\,
  52^\mathrm{m}\, 55^\mathrm{s}.43 \, , \mbox{ Dec.: }{-29}^\circ\,
  27'\, 19''.6$), which has 2 free parameters (the velocity dispersion
  $\sigma_\mathrm{v}$ and the core radius $r_\mathrm{c}$);
\item[NISf] A non-singular isothermal sphere with free coordinates for
  its center, which has 4 free parameters ($\sigma_\mathrm{v}$,
  $r_\mathrm{c}$, and the coordinates of the lens center, $\vec\theta_0$);
\item[NFWwl] A NFW profile \citep[see][]{1996A&A...313..697B} centered
  at the maximum of the weak lensing mass map, which has 2 free
  parameters (the scale radius $r_\mathrm{s}$ and the density
  $\Sigma_0 = \rho_\mathrm{s} r_\mathrm{s}$);
\item[NFWf] A NFW profile with free center, which has 4 parameters
  ($\Sigma_0$, $r_\mathrm{s}$, and $\vec\theta_0$).
\end{description}

\begin{table}[b!]
  \centering
  \begin{tabular}{lcccccc}
    Model & \multicolumn{2}{c}{$\vec\theta_0$ [arcsec]} & 
    $\sigma_\mathrm{v}$ [km s$^{-1}$] & $r_\mathrm{c}$ [arcsec]\\
    \hline
    NISwl & -- & -- & 1365 & 12.82 \\
    NISf  &  $-9.61$ & $+0.10$ & 1185 & 5.91  \\
    \hline
    \hline
    Model & \multicolumn{2}{c}{$\vec\theta_0$ [arcsec]} & 
    $\Sigma_0$ [$\mbox{M}_\odot \, \mbox{pc}^{-2}$] & $r_\mathrm{s}$
    [arcsec] \\
    \hline
    NFWwl & -- & -- & 320 & 108 \\
    NFWf  & $-10.17$ & $1.78$ & 523 & 52 \\    
  \end{tabular}
  \caption{The best fit parameters for the various models.  The center
    $\vec\theta_0$ is written as celestial coordinates $\Delta \alpha$
    and $\Delta \delta$ with respect to the peak of the weak lensing
    map ($\mbox{R.A.: }12^\mathrm{h}\, 52^\mathrm{m}\,
    55^\mathrm{s}.43 \, , \mbox{ Decl.: }{-29}^\circ\, 27'\,
    19''.6$).
    Note that the scale radius $r_\mathrm{s}$ appearing in
    the NFW models is intrinsically different from the core radius
    $r_\mathrm{c}$ of the NIS models.}
  \label{tab:1}
\end{table}

\begin{figure}[!t]
  \begin{center}
    \includegraphics[bb=152 258 452 575, width=\hsize,
    keepaspectratio]{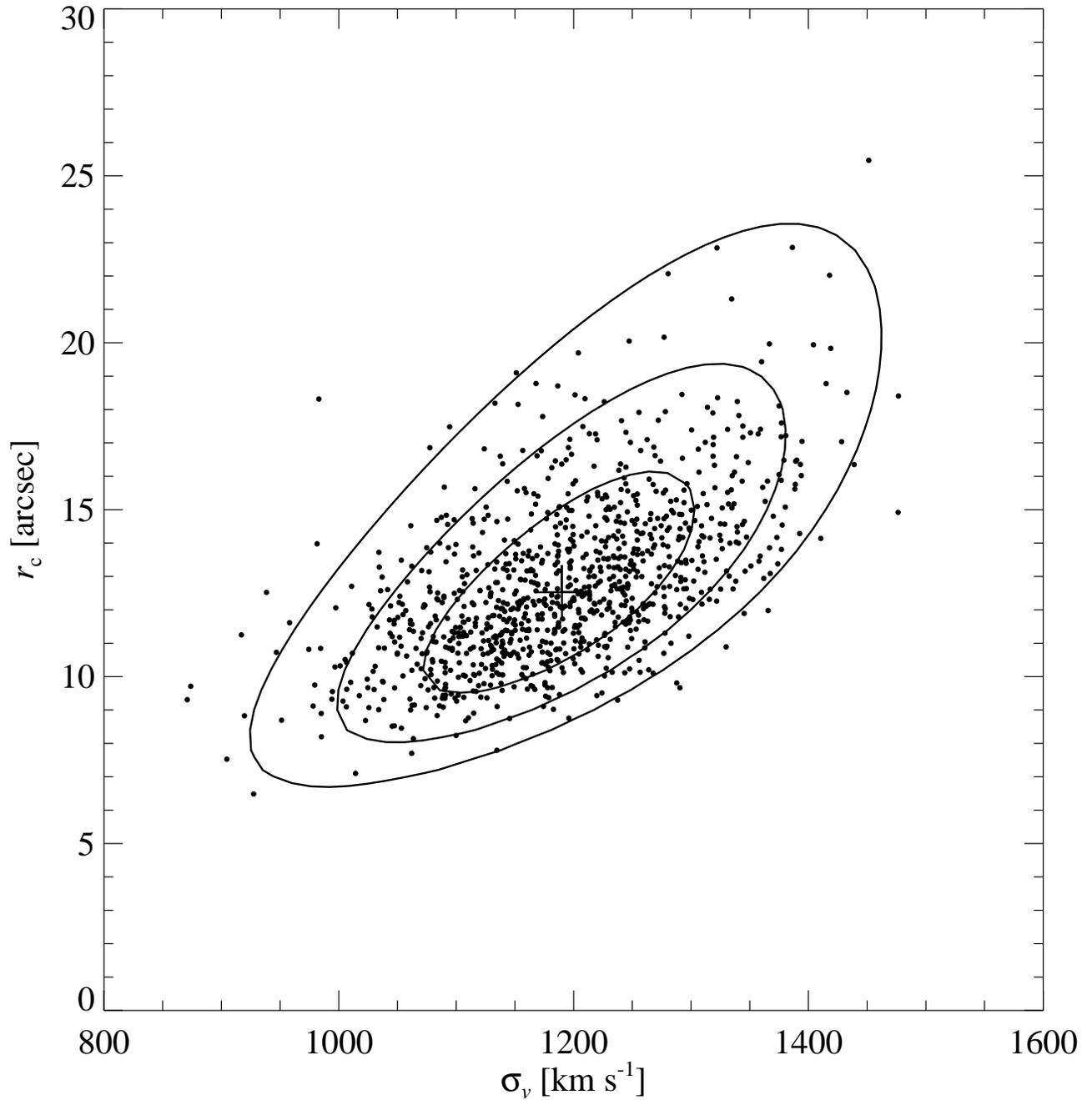}
    \caption{Confidence levels for the non-singular isothermal sphere
      fits NISwl performed on the $i$ band.  The plot shows the
      $68.2\%$, $95.4\%$, and $99.7\%$ confidence regions for the lens
      velocity dispersion $\sigma_\mathrm{v}$ and core radius
      $r_\mathrm{c}$.  The points show the best fit parameters
      obtained by boostrapping $1\,000$ catalogs (see
      text).}
    \label{fig:14}
  \end{center}
\end{figure}

\begin{figure}[!t]
  \begin{center}
    \includegraphics[bb=152 258 461 574, width=\hsize,
    keepaspectratio]{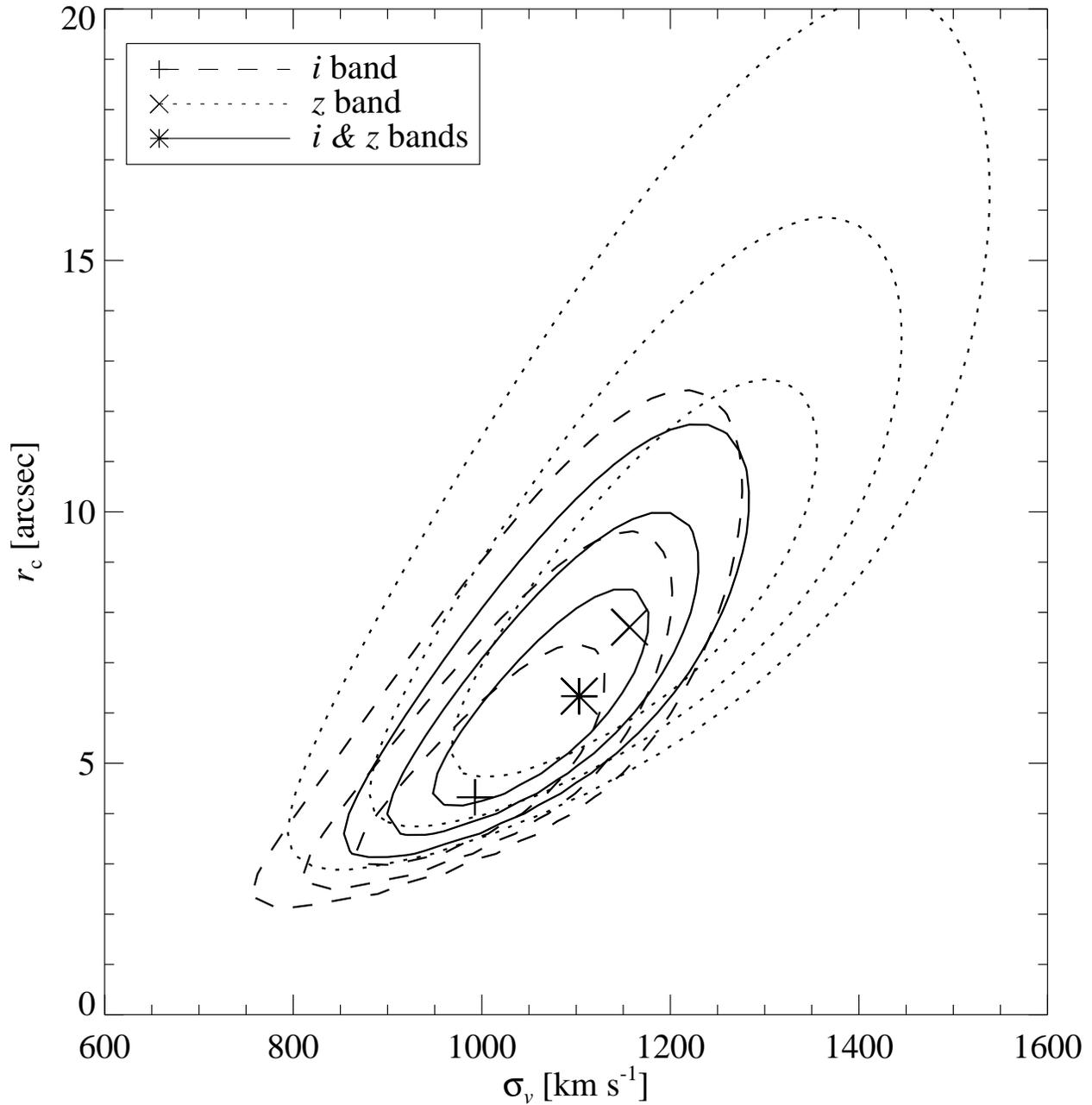}
    \caption{Confidence levels for the non-singular isothermal sphere
      fits NISf.  The plot shows the $68.2\%$, $95.4\%$, and $99.7\%$
      confidence regions for the lens velocity dispersion
      $\sigma_\mathrm{v}$ and core radius $r_\mathrm{c}$ for the fits
      performed using the $i$ catalog, the $z$ catalog, and both.  The
      contours are obtained by marginalizing the results over the lens
      center (see also Fig.~\ref{fig:16}).  The best fit values are
      also marked.}
    \label{fig:15}
  \end{center}
\end{figure}

\begin{figure}[!t]
  \begin{center}
    \includegraphics[bb=139 280 454 554, width=\hsize,
    keepaspectratio]{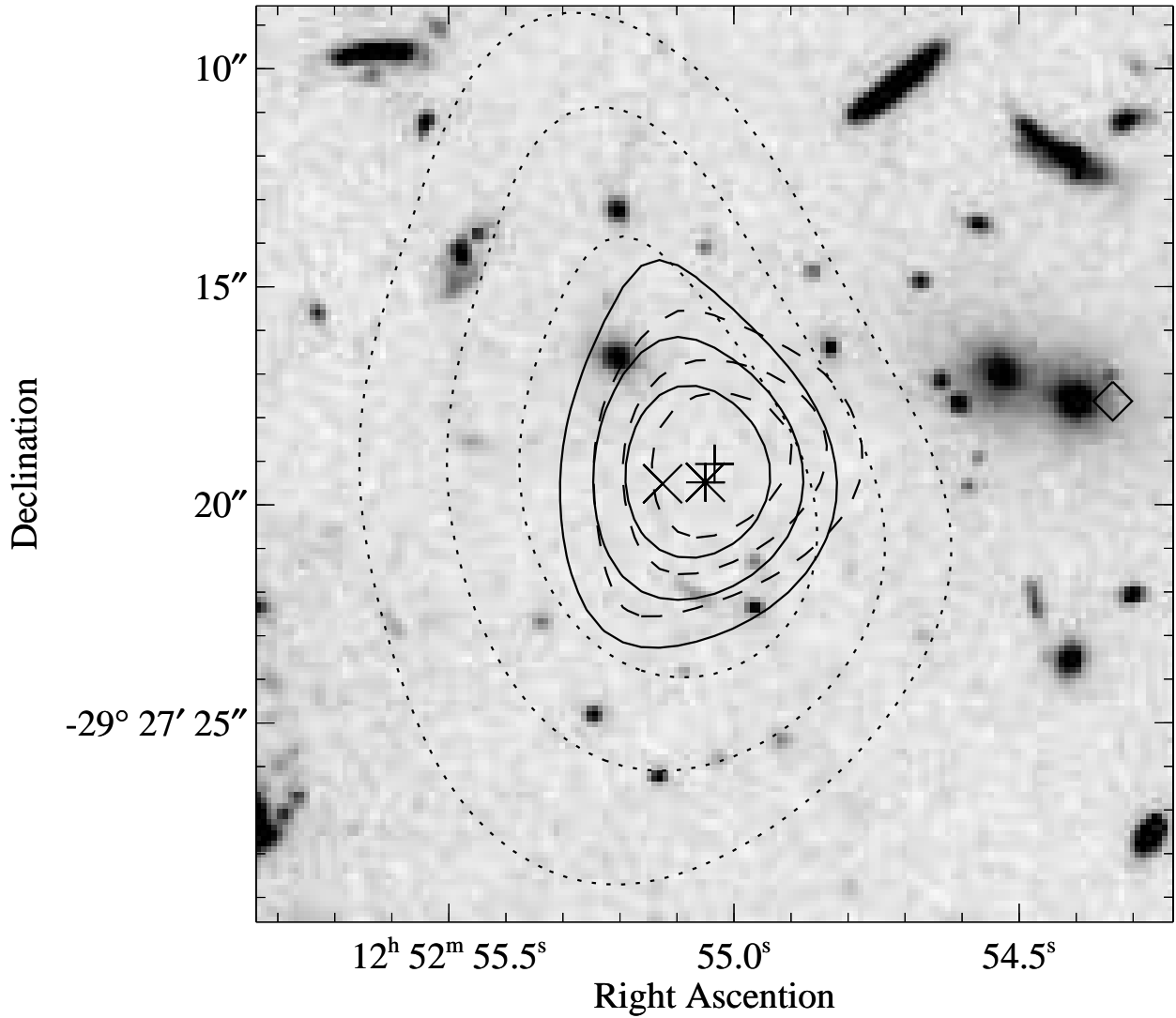}
    \caption{Confidence regions corresponding to the other two
      parameters of the NISf fit, namely the two coordinates of the
      center of the lens, overlaid onto $i$-band image of the field.
      The best fit center appears to have a significant offset with
      respect to the two cD galaxies (center right).  Symbols are as
      in Fig.~\ref{fig:15}; the centroid of the X-ray emission is
      marked with a diamond.  A similar offset is observed for the
      NFWf fit.}
    \label{fig:16}
  \end{center}
\end{figure}

\begin{figure}[!t]
  \begin{center}
    \includegraphics[bb=147 257 461 575, width=\hsize,
    keepaspectratio]{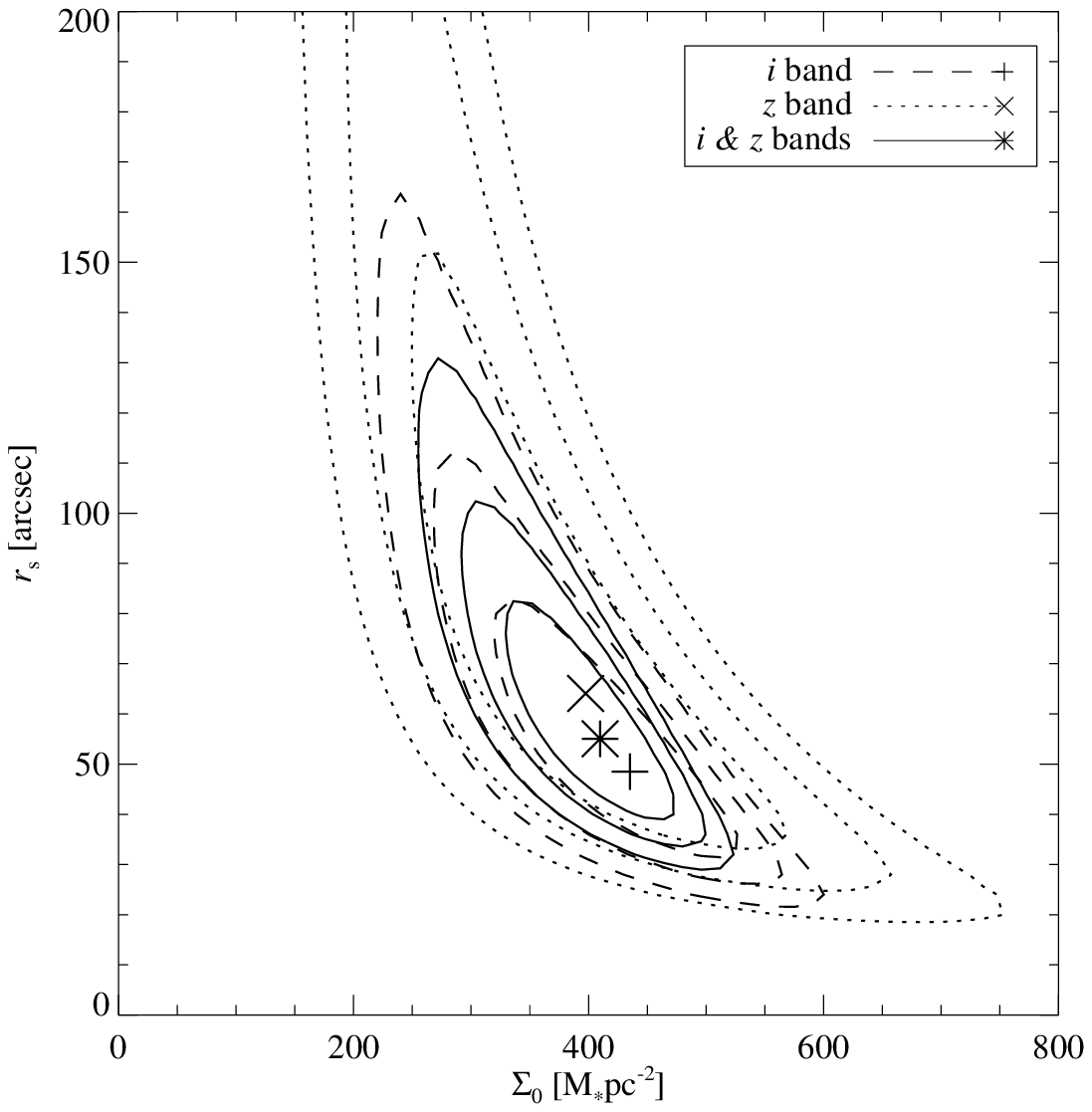}
    \caption{Confidence levels for the NFW spherical fit.
      The plot shows the $68.2\%$, $95.4\%$, and $99.7\%$ confidence
      regions for the lens central density $\Sigma_0$ and
      scale radius $r_\mathrm{s}$ for the fits performed using the $i$
      catalog, the $z$ catalog, and both.}
    \label{fig:17}
  \end{center}
\end{figure}

The results of the best fit parameters are reported in
Table~\ref{tab:1} for both the NIS and the NFW models.  Note that for
both the NIS and the NFW models we obtained mass estimates
inconsistent with the virial estimate, based on a measured galaxy
velocity dispersion $\sigma_\mathrm{v} \simeq 750 \mbox{ km s}^{-1}$
(\citealp{2004AJ....127..230R}; see below for a discussion of this
point).  Results for the NISwl fit performed using $i$-band catalog
are shown in Fig.~\ref{fig:14}, where we plotted the confidence levels
obtained from the likelihood ratio technique \citep[see,
e.g.,][]{Eadie}.  In order to test the reliability of the likelihood
ratio, we generated $1\,000$ catalogs by adding random errors to the
original galaxies ellipticities.  In particular, we added to the shear
estimates of each galaxy a Gaussian error with variance equal to the
estimated shear error.  As shown in Fig.~\ref{fig:14}, the confidence
regions reproduce very well the density of fitted parameters (this is
also confirmed by a quantitative statistical analysis, not reported
here).  Note that, since the confidence regions shown in
Fig.~\ref{fig:14} are in a two-dimensional parameter space, the
probability of having a velocity dispersion smaller than, say, $900
\mbox{ km s}^{-1}$ \textit{regardless\/} of the value of the core
radius $r_\mathrm{c}$ is extremely low (about $1.4\%$).

Figure~\ref{fig:15} shows confidence regions in the
$\sigma_\mathrm{v}$--$r_\mathrm{c}$ slice of the parameter space for
the NISf fit; the three regions correspond to the fit performed on the
$i$, $z$, and $i + z$ catalogs.  A comparison of the confidence
regions for the $i$ and $z$ bands shows that the latter have a slight
offset toward larger velocity dispersions.

The other slice of the parameter space for the NISf fit, namely the
lens position $\vec\theta_0$, is shown in Fig.~\ref{fig:16}, together
with the $i$ band image in the background.  We observe a significant
offset (about $8 \mbox{ arcsec}$ to the East, corresponding to $65
\mbox{ kpc}$ at the redshift of the cluster) of the lens center with
respect to the light, and in particular to the two cD galaxies.
Interesting, a similar offset is observed also when fitting with the
NFWf profile and on the parameter-free mass maps (cf.\ 
Figs.~\ref{fig:6} and \ref{fig:8}).  All these observations seem to
indicate a real offset between the cluster baryonic and dark matter
distributions.  Special care is required to interpret this result
because (i) the center of the parametrized models could be biased
because of the use of simple, radially symmetric profiles; (ii) the
offset on the parameter-free maps is much smaller than the smoothing
length used to build them ($\sigma_\mathrm{W} = 25 \mbox{ arcsec}$);
(iii) the center of the weak lensing mass could be offset because of
the effect of other intervening mass concentrations (see below
Sect.~\ref{sec:effect-aligned-mass}).

A similar analysis was performed for the NFW models.  As an example,
we report here the confidence regions for the
$\Sigma_0$--$r_\mathrm{s}$ slice on the NFWf fit are shown in
Fig.~\ref{fig:17} for the various catalogs.

\begin{figure*}[!t]
  \begin{center}
    \includegraphics[bb=137 258 458 577, width=0.48\hsize,
    keepaspectratio]{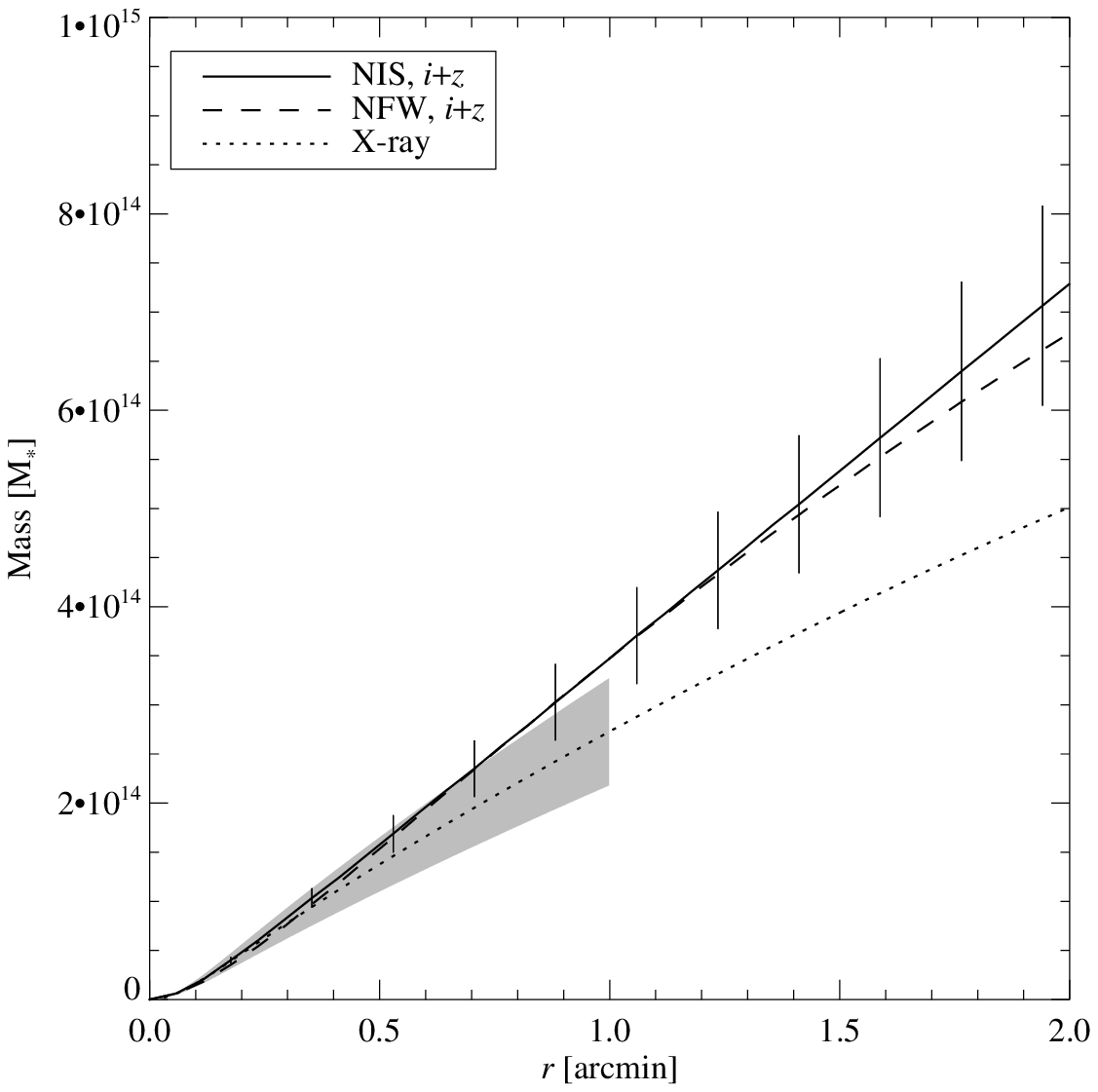}
    \hfill
    \includegraphics[bb=130 258 453 577, width=0.48\hsize,
    keepaspectratio]{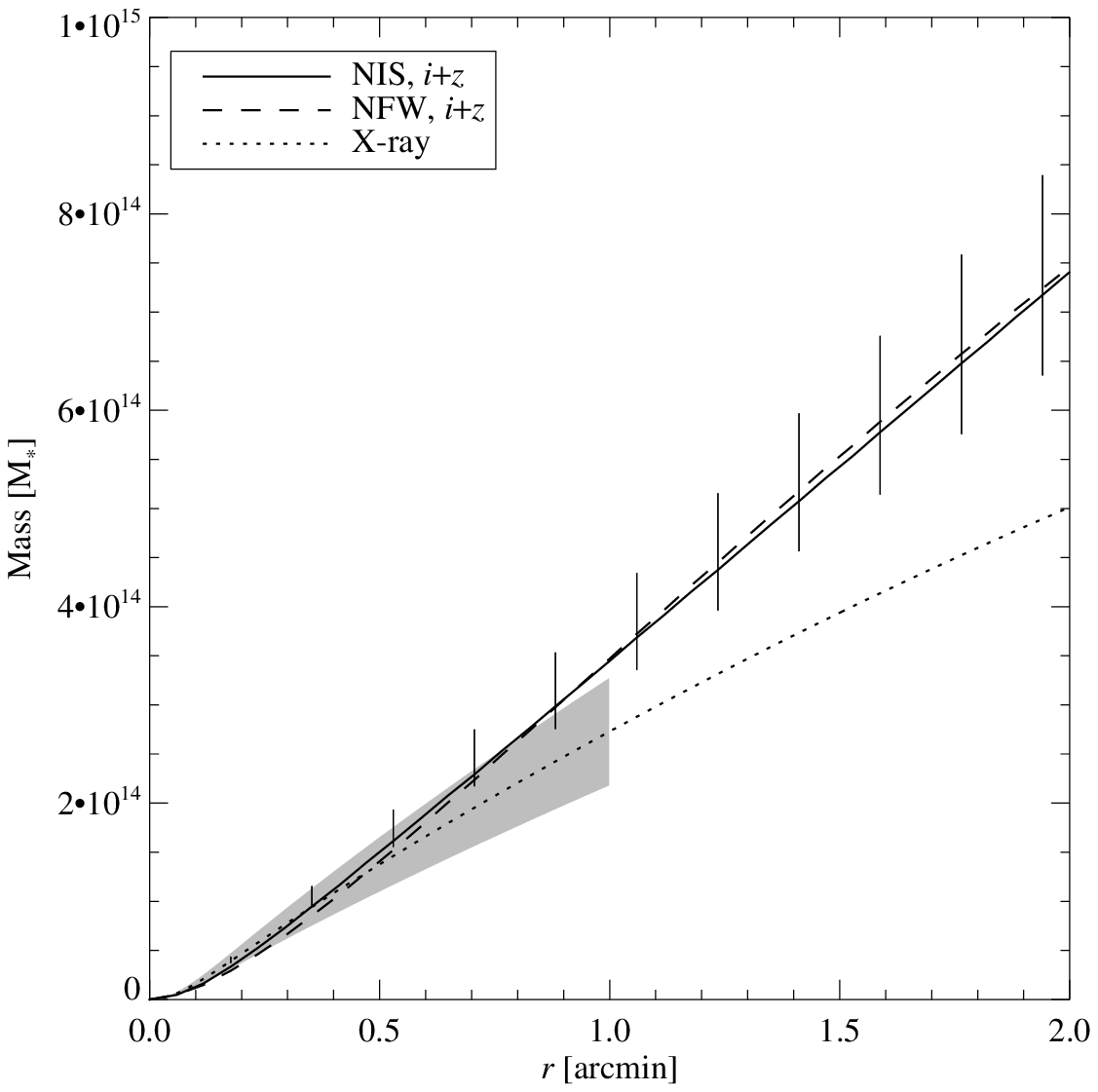}
    \caption{The cumulative radial profiles of the best fit NIS and
      NFW models compared with the projected mass determined from the
      X-ray data for different temperatures.  The gray area denotes
      the region over which X-ray emitting gas is detected ($1 \mbox{
        arcmin} \simeq R_\mathrm{500}$), and its thickness represent
      the estimated 1-$\sigma$ error.  Weak lensing mass profiles are
      derived using the best-fit centers in the left panel, and the
      X-ray center in the right panel.}
    \label{fig:18}
  \end{center}
\end{figure*}

Finally, we show the mass integral radial profiles for the different
fits in Fig.~\ref{fig:18}; errors for the NIS fit are plotted as
vertical segments (errors on the other fits have a similar trend and
are not shown here).  As pointed out in several studies
\citep[e.g.][]{2002A&A...383..118K}, weak lensing techniques are not
very effective in discriminating among different mass models.

We show in Fig.~\ref{fig:18} the X-ray mass measurement, projected
along the line of sight, with its uncertainty (see
\citealp{2004AJ....127..230R} for details).  We note that the X-ray
and lensing mass estimates are consistent within $1 \mbox{ arcmin}$,
which is the radius out to which the X-ray emission can be traced.  In
particular, the two profiles are statistically indistinguishable up to
$\sim 30''$--$40''$; at larger radii, there is a hint for the weak
lensing mass to be slightly larger than the X-ray mass.  In general,
given the difficulties of both mass measurements, the results obtained
are extremely encouraging and provide further evidence that X-ray and
weak lensing mass estimates can be in very good agreement to each
other when high-quality data are used.

Although the X-ray derived mass compares favorably with the weak
lensing analysis, we find useful to describe in the following
paragraphs some sources of errors and biases of both methods.  The
discussion below, though not exhaustive, can be of interests for other
weak lensing analyses of high-redshift clusters.

\subsubsection{Uncertainties on the X-ray mass estimates}
\label{sec:inaccurate-x-ray}

The central panel of fig.~\ref{fig:12} shows that the surface
brightness profile is smooth on the East side, whereas it shows a
discontinuity (possibly a cold front) on the West side, where
presumably a departure from hydrostatic equilibrium occurs.  In order
to evaluate the impact of this anisotropy, we recalculated the $\beta$
model \citep{1978A&A....70..677C, 1976A&A....49..137C} of the cluster
for two sectors located to the East and to the West of the X-ray
centroid (see Fig.~5 of \citealp{2004AJ....127..230R}).  The results
obtained show that there is only a small ($~ 6\%$) increase in the
mass estimate inside $500$--$530 \mbox{ kpc}$ if the X-ray $\beta$
model is derived from the West sector only.

Another source of uncertainty on the X-ray mass estimate is given by
the temperature profile.  Our X-ray mass estimate assumes an
isothermal ICM with a temperature $T=(6.5 \pm 0.5) \mbox{ keV}$ (the
best fit from the \textit{Chandra\/} and \textit{XMM-Newton\/} data).
We investigated the presence of a temperature profile within 60 arcsec
from the X-ray center by using the \textit{Chandra\/} data.  By fixing
the metallicity at $0.4 \times Z_{\odot}$, we measure $kT =
7.4^{+2.1}_{-1.5}, 6.1^{+1.4}_{-1.1}, 5.5^{+3.9}_{-1.9}$ keV in the
radial bins $(0$--$10) \mbox{ arcsec}$, $(10$--$35) \mbox{ arcsec}$
and $(35$--$60) \mbox{ arcsec}$, respectively, with $\sim 300$ net
counts each.  These values suggest indeed that the gas temperature is
higher in the central $35 \mbox{ arcsec}$ than actually assumed for
our mass estimate and decreases outward. A polytropic profile with
index $\gamma = \log T / \log n_{\rm gas} \approx 1.14$ and a central
temperature value $T_0 = (7.6 \pm 2.0) \mbox{ keV}$ provides a good
fit to the observed temperature profile.  Being the mass measurement
directly proportional to $\gamma T_0$ ($\gamma=1$ and $T_0=T$ for the
isothermal case), a direct implication of this negative gradient is
that the mass estimates within $60 \mbox{ arcsec}$ are systematically
higher than the ones obtained under the isothermal assumption, with
values larger by $20\%$ within $20 \mbox{ arcsec}$.  On the other
hand, the two estimates agree within $10\%$ at $R_{500} \simeq 530
\mbox{ kpc}$.  We note however that the systematic effect introduced
by a temperature profile in the X-ray mass measurement is comparable
to its statistical uncertainty, $\sim 35\%$ ($1$-$\sigma$ level).

\subsubsection{Uncertainties on the weak lensing mass}
\label{sec:uncert-weak-lens}

As described above, for such a distant cluster the lensing mass
estimate is \textit{strongly\/} dependent on the assumed redshift
distribution of background galaxies, which in this paper has been
parametrized using the effective redshift $z_\mathrm{eff}$.  As
discussed above, this effect alone introduces an error of $\sim 15\%$
in our case.

A possible source of systematic error could be ascribed to an
inaccurate PSF correction performed by \texttt{Imcat}.  Although this
software has been successfully applied to many ground-based and HST
observations, it has never been tested on ACS data, and to our
knowledge the weak lensing analysis presented in this paper is the
first one carried out with \texttt{Imcat} on ACS images.  This point
will be further investigated on a follow-up paper.

Departure from spherical symmetry and substructures are also a
potential source of biases \citep[e.g.][]{2004MNRAS.350.1038C}.
Although \cluster\ appear to be fairly round in both X-ray and optical
images, the presence of a cold front feature does not exclude
substructures in its mass distribution; therefore, the formal errors
of the various fit parameters should be taken with caution.  This
might also contribute to the observed offset between the X-ray/optical
and the lensing mass centroids.

\subsubsection{Effect of aligned mass concentrations}
\label{sec:effect-aligned-mass}

As pointed out by many authors \citep[e.g.][]{2001ApJ...547..560M,
  2002ApJ...575..640W}, intervening structures observed along the line
of sight of a cluster can have significant effects on its weak lensing
mass estimate.  For massive clusters at low-to-moderate redshift this
effect is typically negligible, since it is unlikely that a structure
along the line of sight be able to significantly perturb the shear
field of the cluster.  However, the situation can be quite different
if the cluster is not very massive or is either at high or at very low
redshift (compared to the average redshift of the background
galaxies).  In these cases, the weak shear field produced by the
cluster can be significantly affected by intervening groups or large
scale structures that happen to be aligned along the line of sight.

\begin{figure}[!t]
  \begin{center}
    \includegraphics[bb=152 316 452 518, width=\hsize, 
    keepaspectratio]{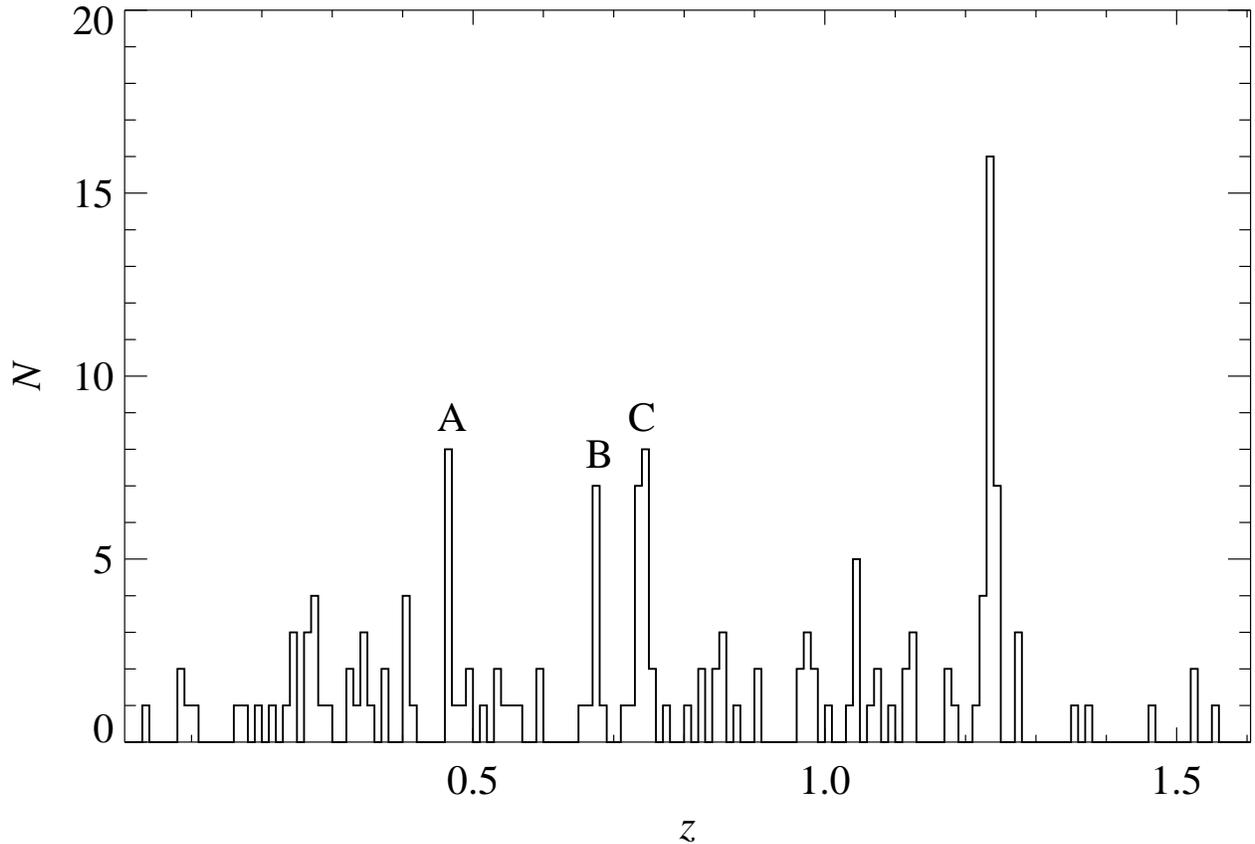}
    \caption{Histogram of the measured spectroscopic redshifts in the
      \cluster\ field.  Three significant foreground groups are
      marked.}
    \label{fig:19}
  \end{center}
\end{figure}

\begin{figure}[!t]
  \begin{center}
    \includegraphics[bb=116 258 456 574, width=\hsize, 
    keepaspectratio]{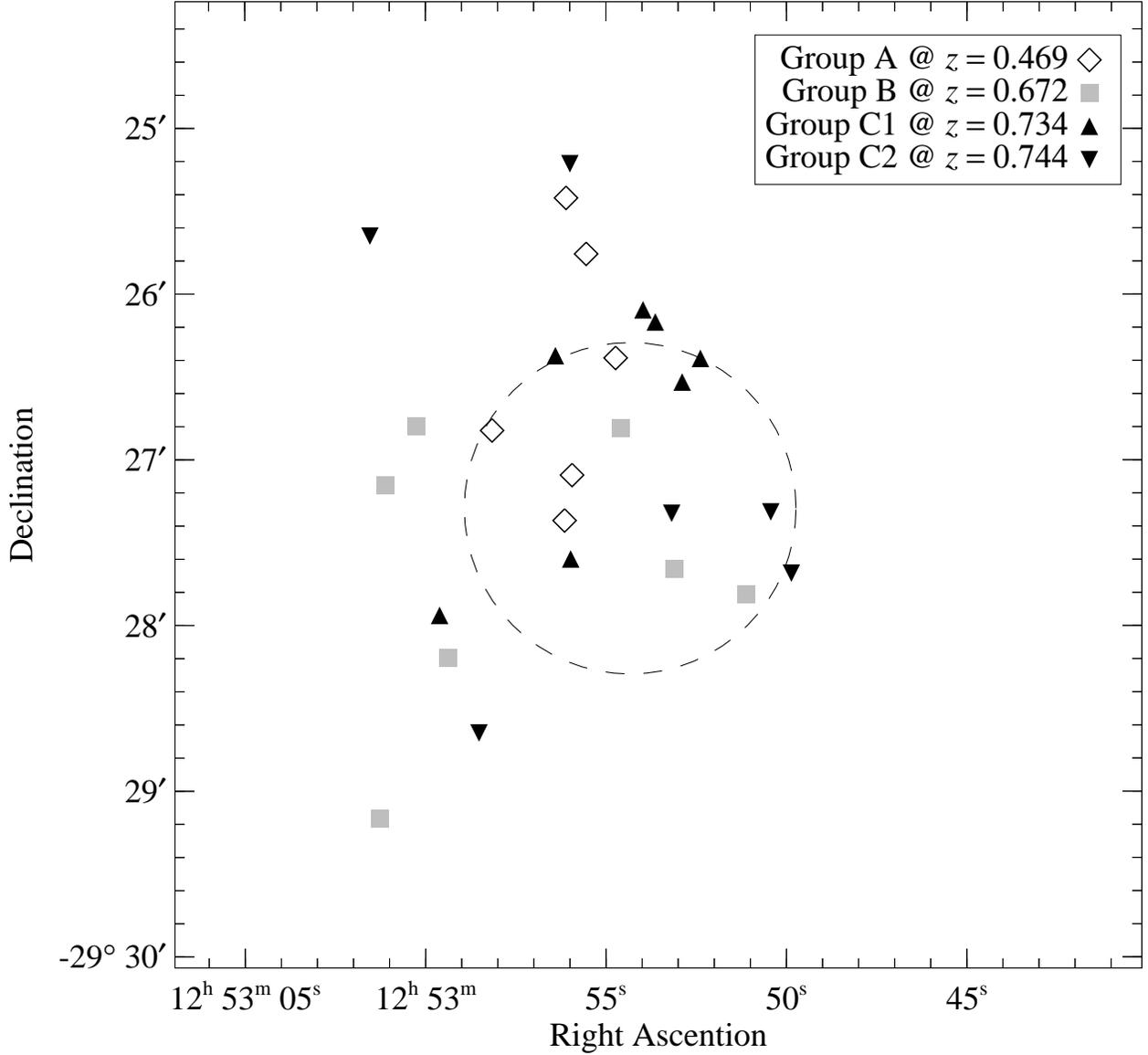}
    \caption{The angular distribution of the galaxies
      belonging to the three redshift peaks identified in
      Fig.~\ref{fig:19}.  The dashed circle denotes the cluster
      X-ray/optical center with $1 \mbox{ Mpc}$ diameter.}
    \label{fig:20}
  \end{center}
\end{figure}

In order to further investigate this point, we have plotted in
Fig.~\ref{fig:19} the distribution of the measured spectroscopic
redshift on the same field.  This figure shows three significant peaks
on the distribution around the redshifts $0.47$, $0.68$, and $0.74$ (a
closer inspection shows that the $z \sim 0.74$ peak in the redshift
histogram is composed of two peaks separated by $\Delta z \simeq
0.01$).  In Fig.~\ref{fig:20} we show the angular distribution of the
galaxies with measured redshift associated with the various peaks,
which appear to correspond to galaxy groups projected along the line
of sight of \cluster.

Because of their lower redshift, these groups can have a
non-negligible impact on our weak lensing mass.  A quantitative
estimate of this effect would require knowledge of individual masses
of the various groups, and is thus difficult at this stage.
Specifically, a mass of $5 \times 10^{12} \mbox {M}_\odot$ for each
group would be responsible for $\sim 15\%$ of the lensing signal.  We
also note (cf.\ Fig.~\ref{fig:16}) that the contribution from these
foreground groups could explain the apparent offset toward the East of
the center of the weak lensing mass.

In conclusion, the lensing mass of high-redshift clusters is likely
biased toward large values because of intervening galaxy groups at
lower redshifts.  With accurate photometric redshifts based on our
multi-wavelength observations, we will attempt to disentangle the weak
lensing effects of the cluster from the ones of the main intervening
masses, a method known as ``weak lensing tomography'' (see
\citealp{2001astro.ph.11605T}; this analysis will be the subject of a
future paper).  Furthermore, strong lensing features observed around
\cluster, which are currently under spectroscopic study, will provide
further constraints on the cluster mass
\citep[see][]{2004astro.ph.10643B}.  The latter is expected to be much
less sensitive to foreground projected masses because of the large
redshifts expected for the arcs.

\subsection{Shear profile}
\label{sec:shear-profile}

\begin{figure*}[!t]
  \begin{center}
    \includegraphics[bb=100 287 496 547, width=0.48\hsize,
    keepaspectratio]{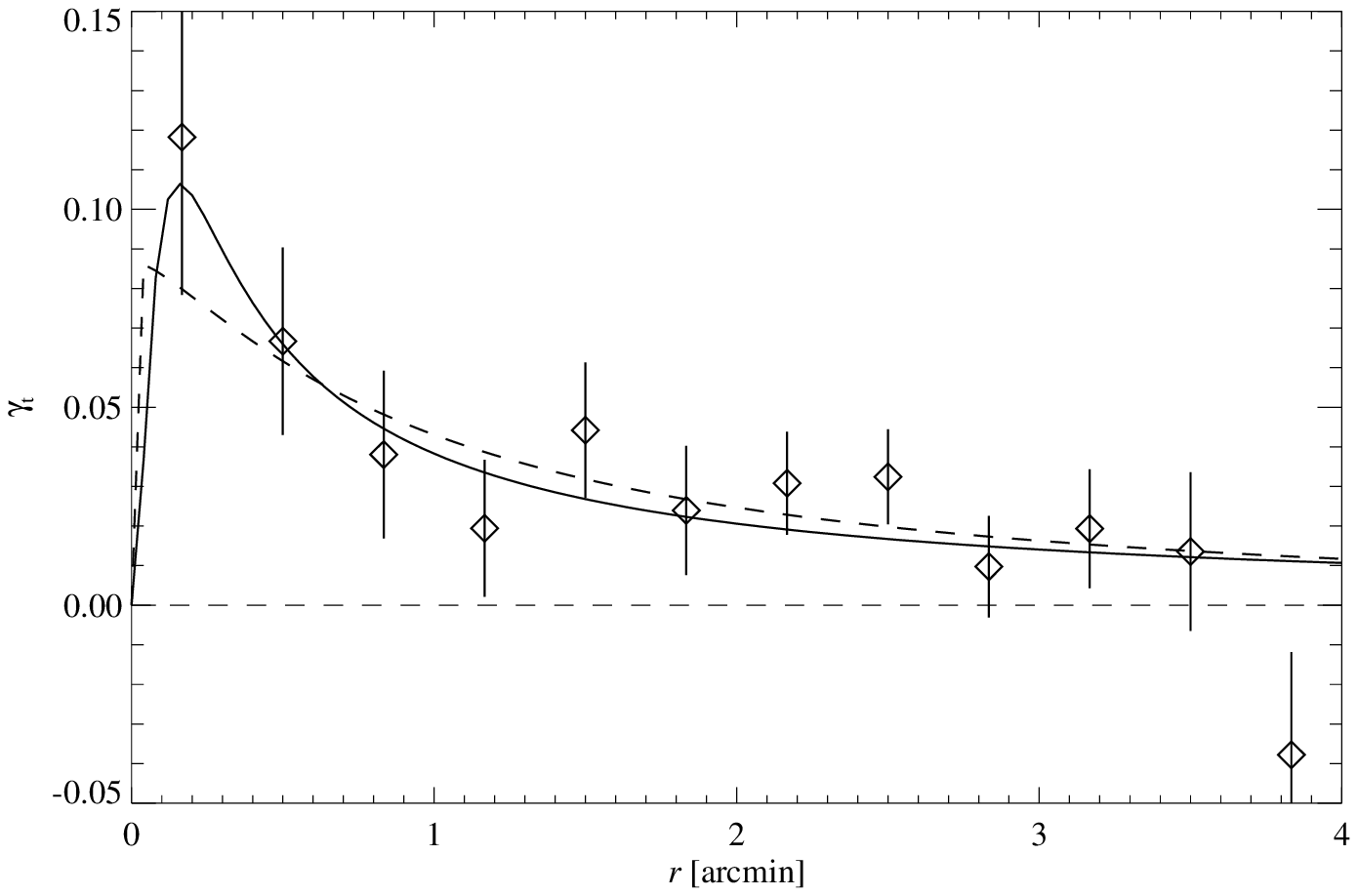} 
    \hfill 
    \includegraphics[bb=100 287 496 547, width=0.48\hsize,
    keepaspectratio]{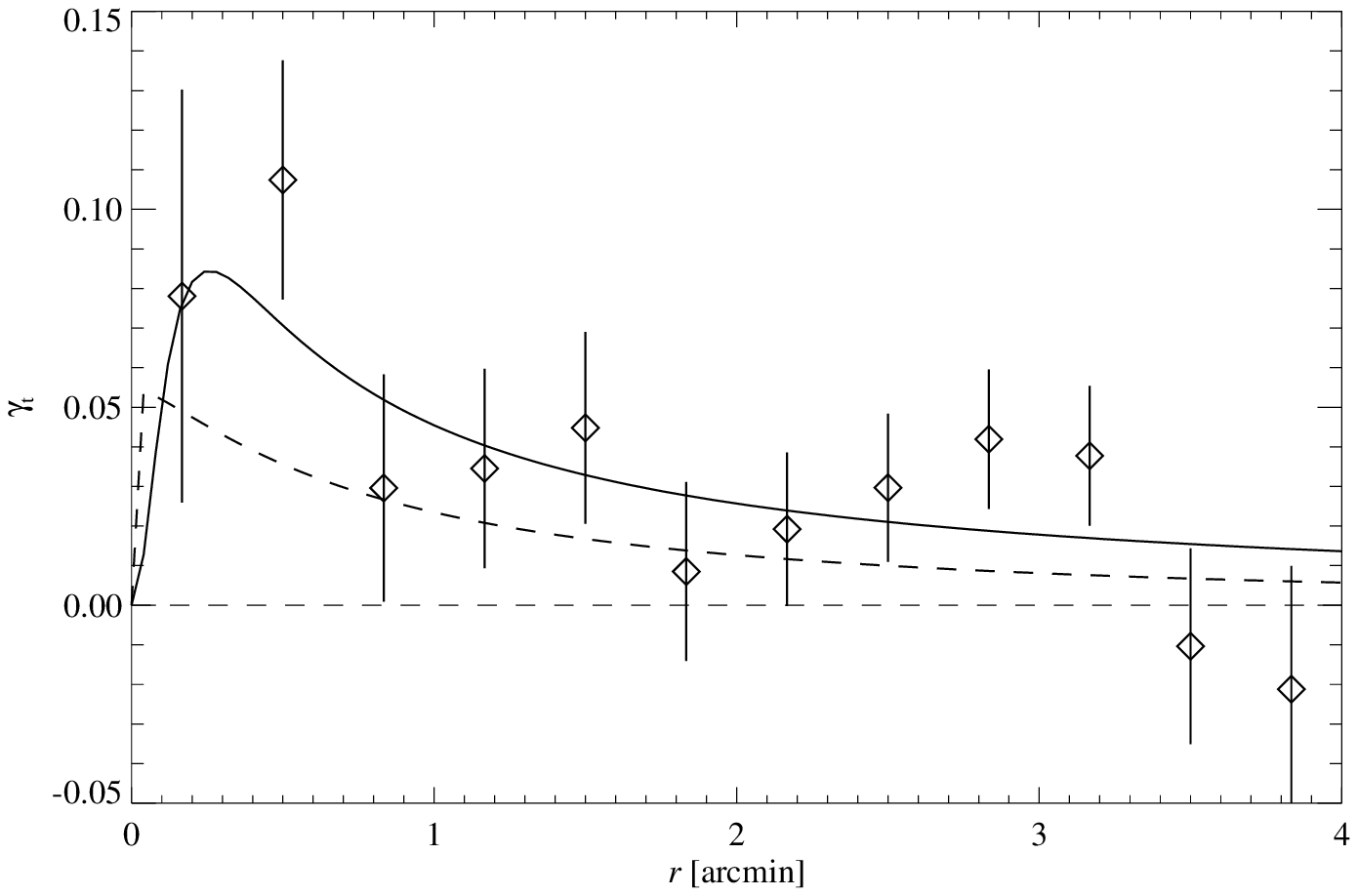}
    \caption{The azimuthally-averaged shear measurements on the $i$
      (left) and $z$ (right) bands and relative errors, with
      overlapped the best-fit NIS (solid line) and NFW (dashed line)
      profiles.  Note that, since individual points in this plot are
      independent, the detection obtained has a high statistical
      significance.}
    \label{fig:21}
  \end{center}
\end{figure*}

\begin{figure*}[!t]
  \begin{center}
    \includegraphics[bb=100 287 496 547, width=0.48\hsize,
    keepaspectratio]{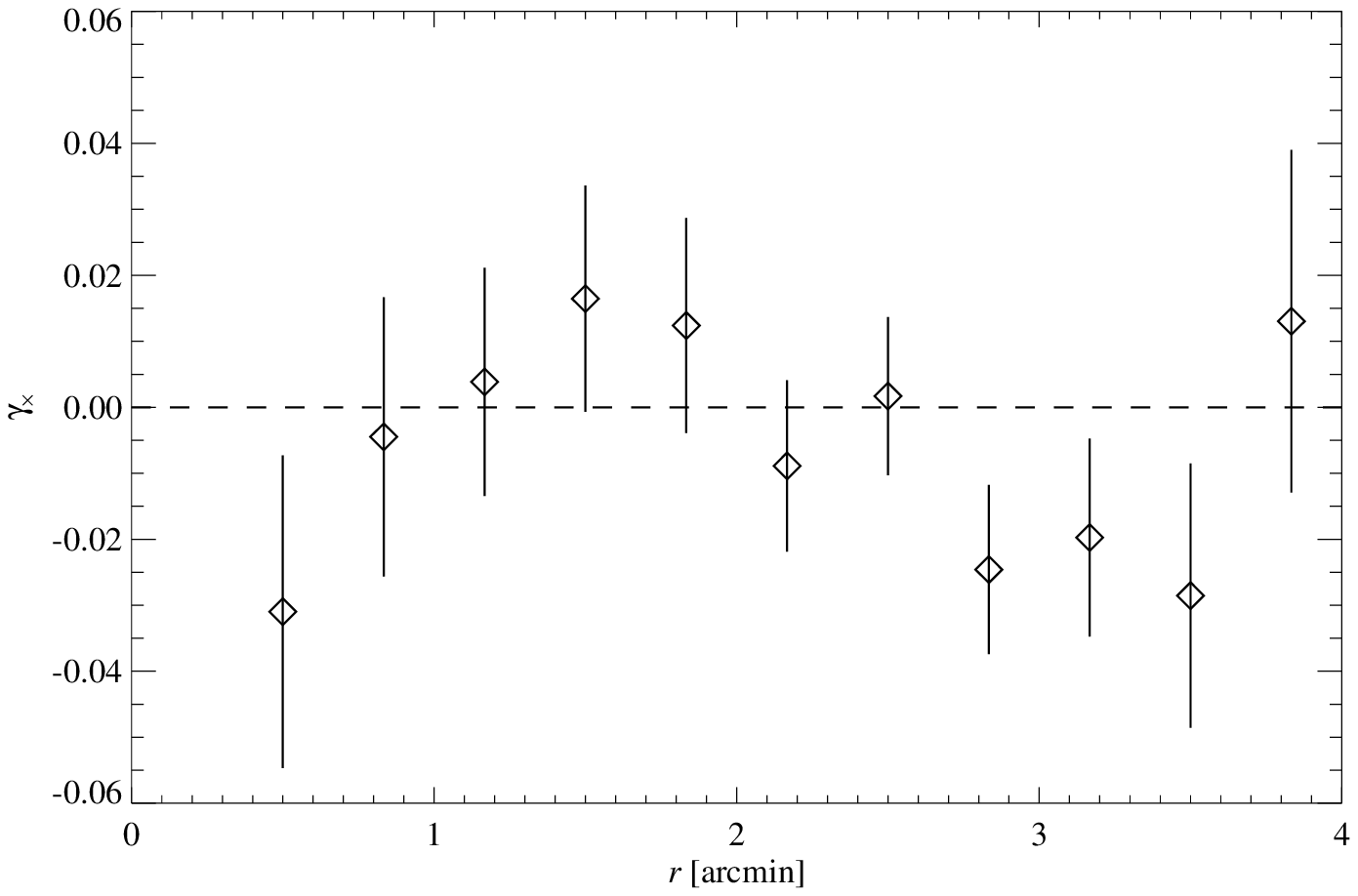}
    \hfill
    \includegraphics[bb=100 287 496 547, width=0.48\hsize,
    keepaspectratio]{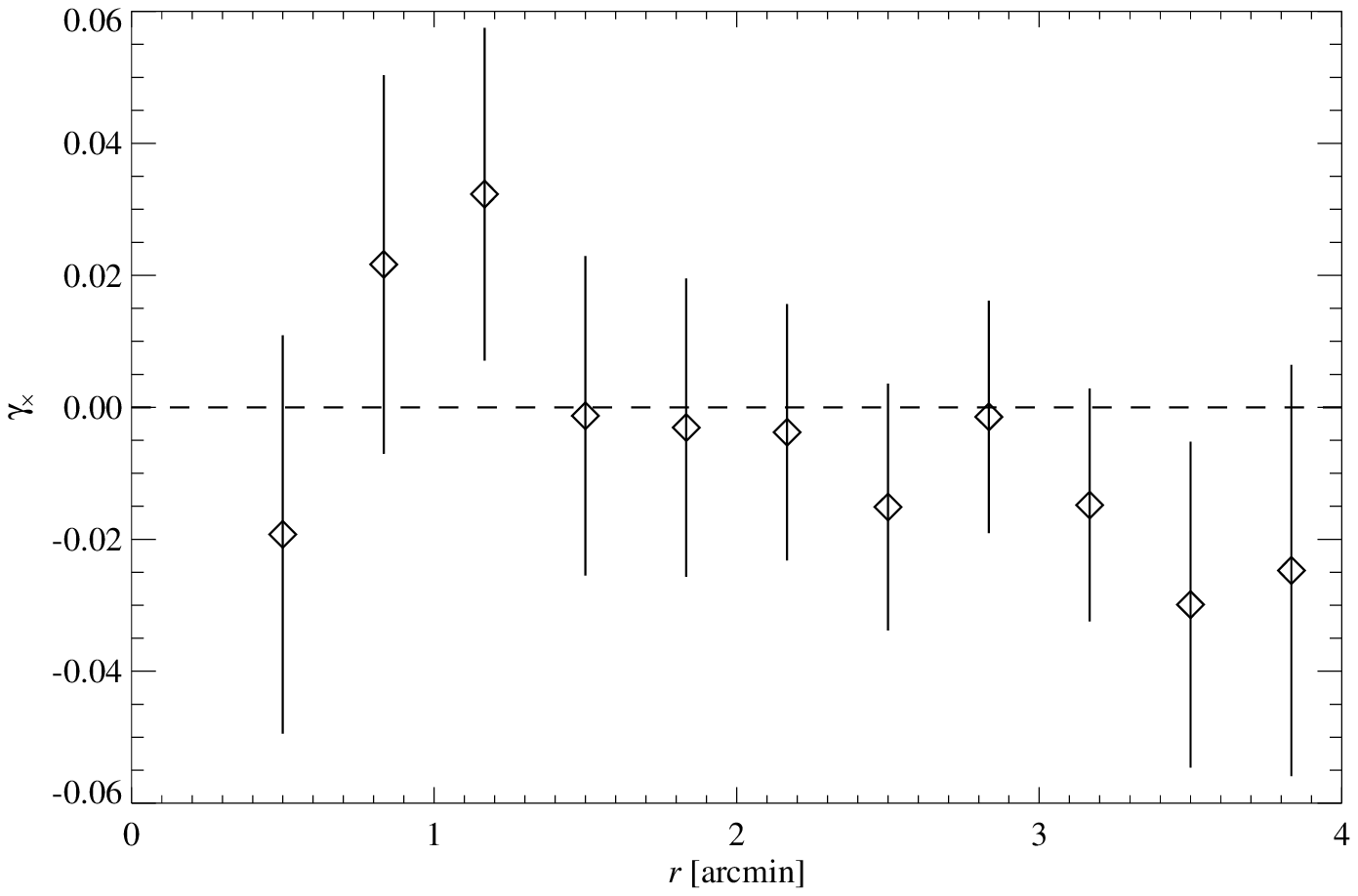}
    \caption{The azimuthally-averaged shear measurements transformed
      according to \eqref{eq:4} on the $i$ (left) and $z$ (right)
      bands and relative errors.  Since lensing is curl-free, we
      measurements are expected to be consistent with zero (cf.\ 
      Fig.~\ref{fig:11}).}
    \label{fig:22}
  \end{center}
\end{figure*}

It is also interesting to investigate the azimuthally averaged shear
profiles, shown in Fig.~\ref{fig:21} for both bands.  We obtained this
plot by computing the average tangential shear on annuli of increasing
radii centered at the peak of the weak lensing mass map.  The errors
reported in this figure were computed from the estimated errors on the
ellipticity of each galaxy.  Since we used different galaxies in
different bins, the various points shown are independent, which
further confirms the high significance of the weak lensing detection
obtained here.  Overplotted is also the tangential shear prediction
from the best-fit NISwl and NFWwl models.

We also evaluated the azimuthally averaged cross shear, obtained by
using the transformation \eqref{eq:4} on the galaxy ellipticities (see
Fig.~\ref{fig:22}).  The null detection obtained in this case is in
agreement with our expectations and shows that there are no
significant systematic errors in our results.

\begin{figure}[!t]
  \begin{center}
    \includegraphics[bb=152 258 453 577, width=\hsize,
    keepaspectratio]{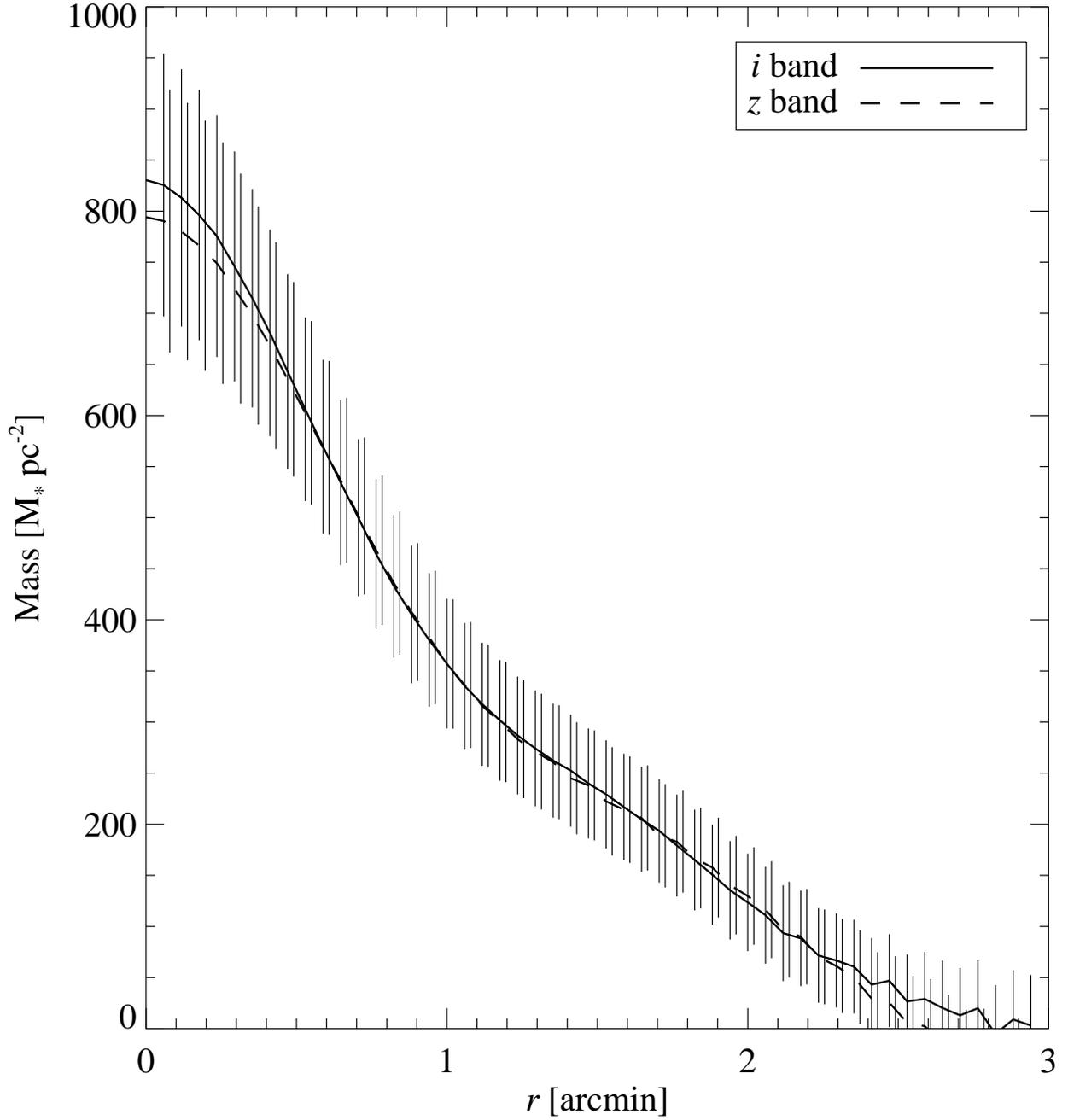}
    \caption{The radial mass profiles obtained from the smoothed weak
      lensing mass maps in the $i$ and $z$ band images (Fig.~\ref{fig:6}).}
    \label{fig:23}
  \end{center}
\end{figure}

Finally, Fig.~\ref{fig:23} shows the radial profiles obtained by
azimuthally averaging the weak lensing mass maps around their peaks in
the $i$ and $z$ band images.  Despite the fact that the galaxies from
which the shear is derived are the same, these two measurements are
not completely dependent because of the contribution of to photometric
noise to the shear error (see discussion at the end of
Sect.~\ref{sec:ellipt-meas}).

\section{Conclusion}
\label{sec:discussion}

In this paper we have discussed the weak lensing analysis of the
massive cluster \Cluster\ at redshift $z = 1.237$, from deep HST/ACS
observations.  The combination of the excellent angular resolution and
sensitivity of ACS, and an accurate measurement of the ellipticities
of background galaxies have allowed us to detect a clear weak lensing
signal in both the $i$ and $z$ SDSS bands.  This result pushes weak
lensing mass reconstructions to unprecedented redshifts and opens the
way to new applications of this technique on the most distant clusters
known to date.

The main results can be summarized as follows:
\begin{itemize}
\item We have detected a $5$-$\sigma$ weak lensing signal in the $i$
  band and a $3$-$\sigma$ signal in the $z$ band; the combination of
  the two shear maps has lead to a $6$-$\sigma$ detection.
\item Several tests based on Monte-Carlo simulations and analytical
  estimates have been performed to ensure that the resulting signal is
  due to lensing and not to systematic effects.
\item The estimate of the differential radial mass profiles in both
  bands has been found to be in excellent agreement with each other.
\item We have fitted the observed galaxy ellipticities with simple
  parametric mass models (NIS and NFW) and have shown that the
  resulting radial profiles are statistically consistent with those
  obtained from the X-ray analysis.
\item The spatial distribution of the dark matter, as inferred from
  the weak lensing map, and the one of the baryons, as traced by the
  X-ray emitting gas and the cluster galaxies, have a similar
  East-West elongation.  However, we have detect an offset ($\sim 8
  \mbox{ arcsec}$) between the centroid of the weak lensing mass map
  and the optical/X-ray centroid.
\item We have discussed the possible sources of errors and biases of
  the X-ray and lensing mass estimates.  We have argued that
  foreground galaxy groups aligned along the line of sight can alter
  the measured shear, thus biasing the cluster mass high, at a level
  which might be detected with deep ACS observations.  The same
  argument could explain the observed offset between the weak lensing
  map and the optical/X-ray centroid.  We have also discussed how
  departures from hydrostatic equilibrium and isothermality can bias
  the X-ray mass measurements toward larger values.
\end{itemize}

\begin{acknowledgements}
  We would like to thank and Giuseppe Bertin, Tim Schrabback, Peter
  Schneider for stimulating discussions and useful suggestions.  ACS
  was developed under NASA contract NAS 5-32865, and this research is
  supported by NASA grant NAG5-7697.  We also acknowledge support from
  NAG5-10176.  We are grateful for an equipment grant from the Sun
  Microsystems, Inc.
\end{acknowledgements}

\appendix

\section{Mass aperture statistics}
\label{sec:mass-apert-stat}

As a further test on the lensing signal, we evaluated the mass
aperture statistics \citep{1996MNRAS.283..837S}.  In particular, for
each point of the field we estimated the quantity
\begin{equation}
  \label{eq:10}
  \hat M_\mathrm{ap}(\vec\theta) = \frac{\sum_{n=1}^N
  w_n g_{\mathrm{t}n}(\vec\theta) Q\bigl( \lvert \vec\theta -
  \vec\theta_n \rvert \bigr)}{\sum_{n=1}^N w_n} \; ,
\end{equation}
where $w_i$ is the weight assigned to the $n$-th galaxy,
$g_{\mathrm{t}n}(\vec\theta)$ is its shear estimate projected
tangentially with respect to the point $\vec\theta$, and $Q$ is given
by
\begin{equation}
  \label{eq:11}
  Q(\theta) = \frac{6}{\pi} \frac{\theta^2}{\theta_0^2} \left( 1 -
  \frac{\theta^2}{\theta_0^2} \right) \; .
\end{equation}
As shown by \citet{1996MNRAS.283..837S}, the quantity $\hat
M_\mathrm{ap}$ defined in Eq.~\eqref{eq:10} is an estimate of
\begin{equation}
  \label{eq:12}
  M_\mathrm{ap}(\vec\theta) = \int \diff^2 \theta' U\bigl( \lvert
  \vec\theta - \theta' \rvert \bigr) \kappa(\vec\theta') \; .
\end{equation}
In other words, $\hat M_\mathrm{ap}$ estimates the convolution of the
lens convergence with a compensated filter $U(\theta)$.  With our
choice for $Q$, we have
\begin{equation}
  \label{eq:13}
  U(\theta) = \frac{9}{\pi \theta_0^2} \left( 1 -
  \frac{\theta^2}{\theta_0^2} \right) \left( \frac{1}{3} -
  \frac{\theta^2}{\theta_0^2} \right) \; .
\end{equation}
The quantity $\theta_0$ sets the filter scale and, in our case, was
chosen to be $\theta_0 = 2'\, 30''$.  This particular choice for
$\theta_0$ maximizes the signal-to-noise ratio and can be justified by
observing that, at the cluster redshift, this scale is $1.3 \mbox{
  Mpc}$, i.e. approximately the cluster size.

We evaluated the mass aperture $M_\mathrm{ap}$ for both the $i$ and
$z$-band catalogs using a filter scale $\theta_0 / 5 = 30 \mbox{
  arcsec}$.  In both bands we observed a prominent peak at the
position of the cluster.  Because of the simple functional form of
$M_\mathrm{ap}$, it is straightforward to evaluate the expected noise
on this quantity and thus to infer the significance of the detection
obtained.  Thus, we were able to confirm the $5$-$\sigma$ significance
in $i$ and the $3$-$\sigma$ in $z$ bands.

\bibliographystyle{apj} 
\bibliography{../../lens-refs.bib}

\end{document}